\documentclass[journal]{IEEEtran}
\ifCLASSINFOpdf
\else
\fi

\usepackage{graphicx}
\usepackage{amsmath,amssymb,amsfonts}
\usepackage{algorithmic}
\usepackage{graphicx}
\usepackage{textcomp}
\usepackage{tabu}
\usepackage{tabularx}
\usepackage{algorithmic}
\usepackage{amsmath}
\usepackage{amsbsy}
\usepackage{amssymb}

\usepackage[linesnumbered, ruled, noend]{algorithm2e}
\usepackage[font=small,labelfont=bf]{caption}

\def\BibTeX{{\rm B\kern-.05em{\sc i\kern-.025em b}\kern-.08em
		T\kern-.1667em\lower.7ex\hbox{E}\kern-.125emX}}
\usepackage{xcolor}
\DeclareMathOperator*{\argmax}{arg\,max}
\usepackage[noadjust]{cite}

\begin{document}
	%
	\font\myfont=cmr12 at 19pt
	\title{{\myfont Data Freshness and Energy-Efficient UAV Navigation Optimization: A Deep Reinforcement Learning Approach}}
	%
	%
	%
	
	%
	%
	
	\author{Sarder~Fakhrul~Abedin,~\IEEEmembership{Student Member,~IEEE,}
		Md.~Shirajum~Munir,~\IEEEmembership{Student Member,~IEEE,}
		Nguyen~H.~Tran,~\IEEEmembership{Senior Member,~IEEE,}
		Zhu~Han,~\IEEEmembership{Fellow,~IEEE,}
		and~Choong~Seon~Hong,~\IEEEmembership{Senior Member,~IEEE,}
		\thanks{Sarder Fakhrul~Abedin, Md. Shirajum Munir, and ~Choong Seon~Hong are with the Department of Computer Science and Engineering, Kyung Hee University, South Korea (E-mail: saab0015@khu.ac.kr, munir@khu.ac.kr, cshong@khu.ac.kr).
			
			Nguyen H. Tran is with the School of Computer Science, The University of Sydney, Sydney, NSW 2006, Australia (E-mail: nguyen.tran@sydney.edu.au)
			
			Zhu Han  is with the Electrical and Computer Engineering Department, University of Houston, Houston, TX 77004, and also with the Department of Computer Science and Engineering, Kyung Hee University, Yongin 17104, South Korea (E-mail: : zhan2@uh.edu).

		}
	}
	\maketitle
	\vspace{-1.5cm}
	\begin{abstract}
		In this paper, we design a navigation policy for multiple unmanned aerial vehicles (UAVs) where mobile base stations (BSs) are deployed to improve the data freshness and connectivity to the Internet of Things (IoT) devices. First, we formulate an energy-efficient trajectory optimization problem in which the objective is to maximize the energy efficiency by optimizing the UAV-BS trajectory policy. We also incorporate different contextual information such as energy and age of information (AoI) constraints to ensure the data freshness at the ground BS. Second, we propose an agile deep reinforcement learning with experience replay model to solve the formulated problem concerning the contextual constraints for the UAV-BS navigation. Moreover, the proposed approach is well-suited for solving the problem, since the state space of the problem is extremely large and finding the best trajectory policy with useful contextual features is too complex for the UAV-BSs. By applying the proposed trained model, an effective real-time trajectory policy for the UAV-BSs captures the observable network states over time. Finally, the simulation results illustrate the proposed approach is $3.6 \%$ and $3.13 \%$ more energy efficient than those of the greedy and baseline deep Q Network (DQN) approaches.
	\end{abstract}
	

	%
	\IEEEpeerreviewmaketitle

	\section{Introduction}
	%
	%
	%
	%
	\setlength\parindent{25pt}The rapid deployment of the Fifth-generation (5G) wireless network sets an unparalleled criteria for high-quality wireless connectivity and services \cite{jung2011cisco}. 
	As a result, conventional cellular networks face enormous challenges to meet the stringent requirements for different 5G application types, such as enhanced mobile broadband (eMBB), ultra-reliable and low-latency communications (URLLC), and massive machine-type communications (mMTC) applications \cite{abedin2018resource}\cite{abedin2018fog}.
	One of potential solutions is to deploy unmanned aerial vehicles (UAV) in the 5G network environment where UAVs serve the network applications and users as aerial base stations (UAV-BS) \cite{zeng2018cellular}.
	Unlike the conventional wireless network infrastructure, UAV-BSs are more agile and capable of not only providing better coverage but also significantly strengthening the network capability to meet the stringent demands of high capacity with wide coverage, and low delay constraints.
	However, the deployment and autonomous navigation of multiple UAV-BSs in 5G networks are still challenging due to the limited energy capacity of UAV-BSs.
	Moreover, in recent years, the concept of edge computing with 5G \cite{8642832} has also emerged to complement the need for a remote cloud environment for enabling computation oriented communications (COC) applications such as virtual and augmented reality (VR and AR) \cite{sukhmani2018edge}, real-time monitoring and surveillance \cite{liu2019uav}.
	As the demand for the upcoming COC applications \cite{letaief2019roadmap} becomes more prevalent, the need for receiving fresh information data update from different futuristic applications \cite{8892984} requires a new metric, age of information (AoI) \cite{liu2018age}, to measure data freshness apart from the traditional performance metrics for the 5G application types.
	
	In case of the UAV-BS navigation, the existing research \cite{pham2018autonomous,pham2018cooperative,sun2016path,walker2019deep,zhou2015multi,ahmed2016energy} mainly focuses on the path planning and placement of UAV-BSs in the network along with communication and energy constraints.
	However, to incorporate the computation oriented applications at the edge network requires an agile UAV-BS navigation that not only enhances the energy efficiency \cite{7057878} but also ensures the up-to-date data delivery on time for seamless operation of COC applications at the edge computing platform.
	
	Under the above circumstances, we focus on optimizing the UAV navigation considering energy efficiency and the AoI context in the 5G enabled edge computing environment.
	The main contributions of the paper are summarized as follows,
	\begin{itemize}
		\item First, we formulate an energy-efficient UAV-BS navigation optimization problem in the edge computing network under the contextual constraints such as energy, navigation, and AoI metric, and then we show that the formulated problem is NP-hard.
		\item  Second, we employ a deep reinforcement learning technique, deep Q-network with \emph{experience replay memory}, which can achieve energy-efficient UAV navigation under the contextual constraints.
		With the proposed trained model, an effective real-time trajectory policy for the UAV-BSs can be obtained that captures the observable network states over time.
		As a result, we design the state, observation, action space and reward explicitly for the proposed deep Q network (DQN) with experience replay which can effectively solve the trajectory optimization problem under the AoI and energy-efficiency constraints.   
		Also, unlike the traditional deep Q network, the proposed model utilizes the benefit of experience replay memory to obtain the optimal trajectory policy while coordinating multiple UAV-BS locations.
		
		\item Finally, we perform an extensive experimental analysis to evaluate the performance of the proposed approach.
		We also conduct an extensive simulation analysis to find the appropriate system parameters to empower the learning model with the proper discount factor and AoI performance metric threshold.
		The results show that the navigation policy that is obtained by applying the proposed DQN with experience replay achieves significant energy efficiency and data freshness compared to the baseline approaches.
	\end{itemize}
	
	The remainder of the paper is organized as follows.
	In Section II, we present an extensive literature review based on the current research. In Sections III and IV, we present the system model and problem formulation, respectively. 
	Section V explains in detail how we solve the proposed optimization problem with deep Q-learning with experience replay.
	In Section VI, we present the simulation analysis to validate the performance and efficiency of our proposed approach for UAV-BS navigation. 
	Finally, in Section VII we conclude the discussion.

	
	%
	
	\section{Literature Review}
	\subsection{UAV Navigation}
	In \cite{wang2019autonomous}, the authors proposed an online deep reinforcement learning approach to enable UAV navigation in a large-scale complex environment.
	The problem formulation for the UAV navigation is based on the partially observable Markov decision process (POMDP) where the authors proposed the actor-critic framework to solve the problem.
	In \cite{huang2019deep}, the authors focused on capturing the UAV motion while planning the trajectory for UAV navigation through massive multiple-input-multiple-output (MIMO) technique.
	The UAV agent makes the navigation decision on the basis of the received signal strengths which are used to train the proposed DQN.
	A trajectory planning method for UAVs in urban environments is proposed in \cite{ragothaman2019multipath}, where the authors considered the UAV's three-dimensional (3-D) environment map to enable navigation to fuse global navigation satellite (GNSS) signals with ambient cellular signals of opportunity.
	A novel Deep Reinforcement Learning (DRL) algorithm is proposed in \cite{zeng2019navigation} for non-holonomic robots with continuous control in an unknown dynamic environment with moving obstacles.
	A distributed sense-and-send protocol to coordinate the UAVs for sensing and transmission is proposed in \cite{hu2019reinforcement}. 
	A reinforcement learning technique is therefore applied to solve some of the UAV's key problems related to trajectory control and wireless resource management.
	\subsection{Energy Efficiency}
	In \cite{zeng2019energy}, the authors strive to reduce the total energy consumption of UAVs, including both power propulsion and communication-related energy, while meeting the requirement of multiple ground node (GN) communication throughput.
	The problem formulation considers the issue of energy minimization by jointly optimizing the allocation of UAV trajectory and contact time between GNs, as well as the overall mission completion time. Using the successive convex approximation (SCA) technique, an effective iterative algorithm is proposed to update the UAV trajectory and contact time allocation at the same time at each iteration, which may converge to a solution that satisfies the Karush-Kuhn-Tucker (KKT) conditions.
	The goal of the work in \cite{dong2019energy} is to reduce the propulsion energy consumption of the UAV while meeting the requirement of throughput by optimizing the trajectory of the running track. To transform the problem into a discrete counterpart, a variable discretization approach is used and later, the problem is transformed into a problem of convex optimization where the proposed method can obtain a locally optimal solution.
	In order to address the critical issue of insufficient on-board UAV energy and CE transmission energy, the authors in \cite{yang2019energy} focused on maximizing energy efficiency (EE) by jointly optimizing the scheduling of the backscatter devices (BD), the power reflection coefficients of the BDs, the transmission power of the CEs and the trajectory of the UAV.
	Moreover, the authors considered the BD's throughput and other realistic constraints for the problem formulation.
	In \cite{hu2019uav}, the authors considered minimizing UAV and user equipment (UE) weighted sum energy consumption subject to task constraints, information-causality constraints, bandwidth allocation constraints, and trajectory constraints of the UAV.
	The UAV-BSs energy-efficient repositioning trajectories are designed in \cite{peng2019predictive } using the Kuhn-Munkres based algorithm, where an Echo State Network based algorithm enables user equipment (UEs) to predict future trajectories. 
	\subsection{Age of Information (AoI)}
	In \cite{jia2019age}, the authors proposed a dynamic programming approach to study a problem of UAV path planning and data acquisition with the concept of AoI metric.
	In the proposed approach, the authors jointly considered the selection of data acquisition mode, energy consumption at each node, and age evolution of the information collected by the UAVs.
	The UAV flight trajectory and status update packet scheduling are jointly configured in \cite{abd2019deep} to achieve the required weighted sum for the age-of-information (AoI) values of various processes at the UAV, referred to as weighted sum-AoI. A deep reinforcement learning (RL) algorithm is proposed to achieve the optimal policy that minimizes the weighted sum-AoI, called the age-optimal strategy.
	The authors suggested a combination of ground sensor nodes (SN) and trajectory planning strategy in \cite{tong2019uav} to strike a balance between the upload time of the SNs and the flight time of the UAVs using dynamic programming in different scenarios.
	In \cite{du2019joint}, the authors sought to minimize UAV's total energy consumption by jointly optimizing the association of the Internet of Things Devices (IoTD), the allocation of computer resources, the UAV hovering time, the wireless power duration and the IoTD service sequence.
	The authors proposed a UAV trajectory planning model for data collection in \cite{li2019minimizing }, where the objective is to minimize expired data packets across the entire sensor network system.
	The authors also simplified the original problem into a min-max-AoI-optimal path scheme due to complex constraints and proposed a reinforcement learning-based strategy for the solution.
	
	Unlike the existing works, in this paper, we design the trajectory policy of UAV-BSs considering the trade-off between the energy-efficiency and AoI metric along with other trajectory constraints for the computation oriented communications (COC) applications.
	Moreover, for incorporating the fresh information update by the UAV-BSs, we consider the high frequency Millimeter Wave (mmWave) wireless spectrum for enabling the backhaul communication between the UAV-BSs and ground BS.  
	
	\section{System Model}
\begin{figure}
		\centering
		\captionsetup{justification=centering}
		\includegraphics[scale=0.30]{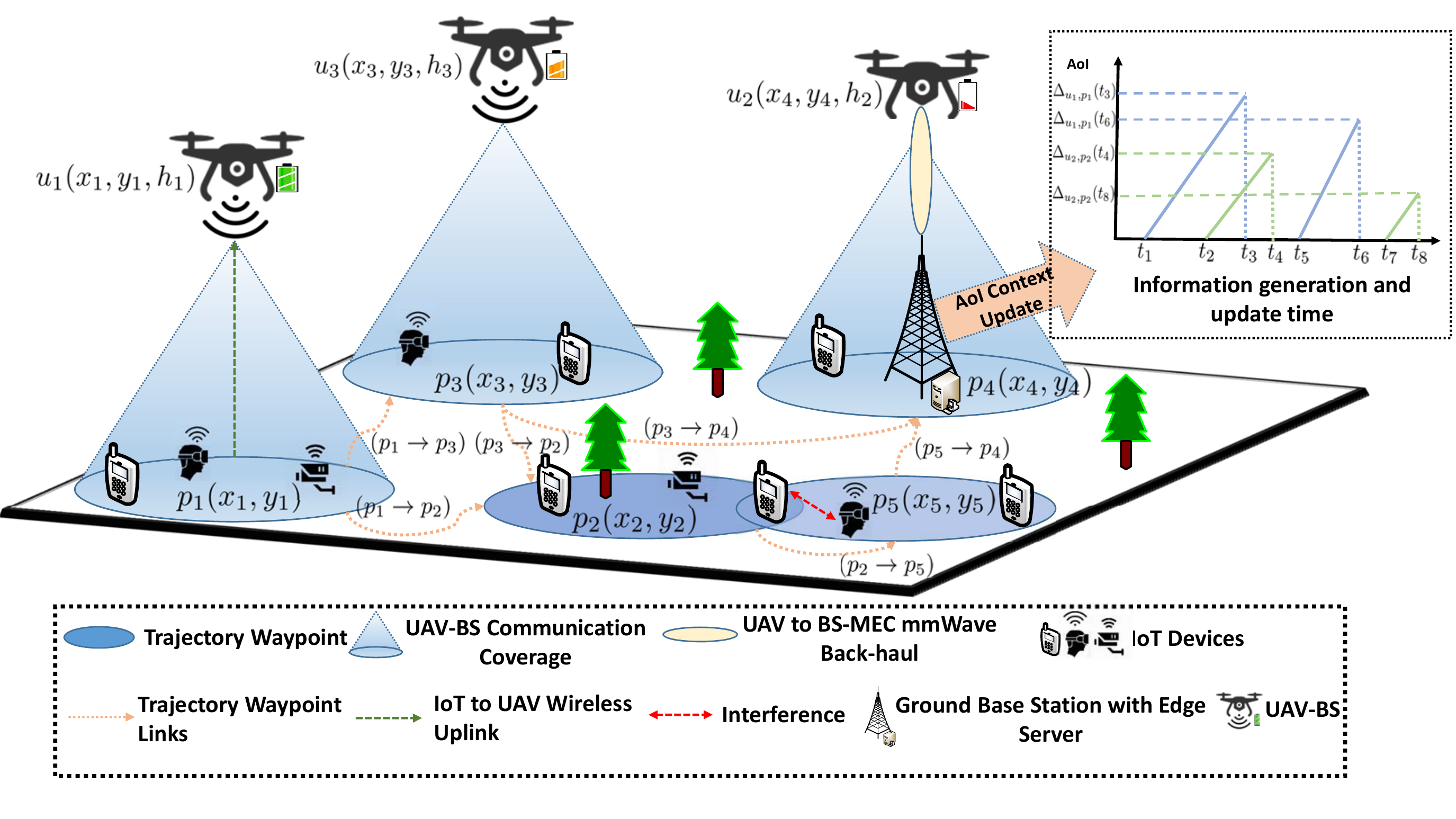}
		\caption{System Model for Heterogeneous Unmanned Aerial Networks with Edge Computing }
		\label{fig1}
		\vspace{-0.2in}	
	\end{figure}

	In Fig. \ref{fig1}, we consider a set of given trajectory points, $\mathcal{P} = \{1,2,\cdots, P\}$.
	The trajectory points in $\mathcal{P}$ are covered by a set of battery-powered UAV-BS, $\mathcal{U} = \{1,2,\cdots,U\}$, which act as relay for a set of IoT devices, $\mathcal{I} = \{1,2,\cdots,I\}$.
	For simplicity, we consider the set of IoT devices are randomly located at different trajectory points.
	In this paper, we also consider a single ground station (i.e., ground base station) $b$ which is equipped with the multi-access edge computing (MEC) server (BS-MEC) and acts as the information fusion center that receives information updates from the IoT devices through the UAV-BS relays to support the computation oriented communications applications.
	Therefore, we assume that the BS-MEC is a point in the network where various communication resources are available to achieve a certain computational accuracy where the timely information updates from the different network sources are essential. 
	Moreover, the BS-MEC $b$ is also considered as a trajectory point within the set $\mathcal{P}$ where the set of neighboring trajectory points of $b$ (exclude itself) is denoted by $\mathcal{P}^{b} = \{ p \in \mathcal{P}:(b,p) \in \digamma\}$ where $\digamma\subseteq \mathcal{P}\times \mathcal{P}$ is the set of trajectory links between the trajectory points.
	The UAV-BS $u \in \mathcal{U}$ traverses within different trajectory points in $\mathcal{P}$ over a finite observation time $T$.
	As UAV-BS $u \in \mathcal{U}$ travels with a trajectory $p \in \mathcal{P}$, it gathers the information data packets from the active IoT device $i \in \mathcal{I}$ located near the trajectory point $p \in \mathcal{P}$ using the \textit{uplink} communication channel.
	Moreover, UAV-BSs $u \in \mathcal{U}$ uses the backhaul communication link when the BS-MEC $b$ is within the transmission range of UAV-BS $u \in \mathcal{U}$.
	As a result, BS-MEC can receive a fresh information update from different trajectory points and calculate the AoI metric.
	We assume that the BS-MEC is equipped with an array of mmWave directional antennas and provides a dedicated mmWave spectrum for back-haul communication for the UAV-BS.
	Moreover, the IoT devices and the UAV-BS are also equipped with directional antennas so that the IoT devices can transmit information updates to the UAV-BS using the non-mmWave spectrum.
	In this paper, we limit the scope by focusing on the deployment, communication, and navigation of the UAV-BSs $u \in \mathcal{U}$ for the data relaying from different sources $i \in \mathcal{I}$ at trajectory points $p \in \mathcal{P}$ to the ground BS $b$.
	Also, for the user association between the IoT devices and the corresponding UAV-BSs at different trajectory points, we apply the default max-signal-to-interference-plus-noise-ratio (SINR) \cite{6497017} based approach. 
	\subsection{IoT-to-UAV-BS Communication Model}
	
	At the trajectory point $p \in \mathcal{P}$, the air-to-ground path loss probability of the UAV-BS $u \in \mathcal{U}$ with IoT devices $ i \in \mathcal{I}$ is calculated as \cite{kalantari2016number},
	\begin{equation}\label{uav_path_loss_probability}
		\zeta_{i,p}^{u} = \left\{
		\begin{array}{ll}
			\frac{1}{1+\alpha \exp(-\hat{\alpha}(\frac{180}{\pi} \Theta_{u} -\alpha))},\; \text{LoS channel,}\\
			1- \bigg[\frac{1}{1+\alpha \exp(-\hat{\alpha}(\frac{180}{\pi} \Theta_{u} -\alpha))}\bigg],\; \text{NLoS channel}.
		\end{array}
		\right.
	\end{equation}
	Here $\alpha$ and $\hat{\alpha}$ are the environment dependent constants for the LoS and NLoS channels, respectively,  where $\Theta_{u}$ is the elevation angle of UAV-BS $u \in \mathcal{U}$.
	The intuition of calculating the  air-to-ground LoS and NLoS path loss probabilities using \eqref{uav_path_loss_probability} is that, in urban/sub-urban environment, the uplink communication link between the the UAV-BSs and the IoT devices may be hindered (i.e., multi-path fading) by the surrounding obstacles (e.g., buildings) unlike IoT devices deployed in the rural environment.
	In addition, the path loss in decibel (dB) is calculated as \cite{al2014optimal},
	\begin{equation}\label{uav_path_loss_model}
		P_{i,p}^{u} = \left\{
		\begin{array}{ll}
			20 \log(\frac{4 \pi f_{c} \delta_{i,p}^{u}}{c}) + \epsilon,\; \text{LoS channel,}\\
			20 \log(\frac{4 \pi f_{c} \delta_{i,p}^{u}}{c}) + \bar{\epsilon},\; \text{NLoS channel}.
		\end{array}
		\right.
	\end{equation} 
	Here $f_{c}$ is the uplink channel frequency, and $\epsilon$ and $\bar{\epsilon}$ are the attenuation factors for the LoS and NLoS channels, respectively.
	Using \eqref{uav_path_loss_model}, the received signal power from the IoT device $i \in \mathcal{I}$ at trajectory point $p \in \mathcal{P}$ to UAV-BS $u \in \mathcal{U}$ is calculated as,
	\begin{equation}\label{uav_received_power}
		\hat{P}_{i,p}^{u} = \frac{\bar{P}_{i,p}^{u}}{P_{i,p}^{u}},
	\end{equation} 
	where $\bar{P}_{i,p}^{u}$ is the transmit power of IoT device $i \in \mathcal{I}$ for offloading the data to UAV-BS $u \in \mathcal{U}$.
	At time $t$, the received SINR for UAV-BS $u \in \mathcal{U}$ with IoT device $i \in \mathcal{I}$ at trajectory point $p \in \mathcal{P}$ is calculated as,
	\begin{equation}\label{uav_iot_sinr}
		\gamma_{i,p}^{u}(t) = \frac{\hat{P}_{i,p}^{u}(10^{\frac{\zeta_{i,p}^{u}}{10}})^{-1}}{I_{i,p}^{u}+\sigma^{2}}.
	\end{equation}
	Here $I_{i,p}^{u} = \sum_{p^{\prime} \in \mathcal{P}} \sum_{u^{\prime} \in \mathcal{U}} \sum_{i^{\prime} \in \mathcal{I}} \hat{P}_{i^{\prime},p^{\prime}}^{u^{\prime}}(10^{\frac{\zeta_{i^{\prime},p^{\prime}}^{u^{\prime}}}{10}})^{-1}$  is the received interference of UAV-BS $ u \in \mathcal{U}$ from the other UAV-BSs $ u^{\prime} \in \mathcal{U}, u \neq u^{\prime}$ which is serving IoT $i^{\prime} \in \mathcal{I}, i \neq i^{\prime}$ that is located in different neighboring and overlapping trajectory points $p^{\prime} \in \mathcal{P}, p \neq p^{\prime}$  and $\sigma^{2}$ is the noise power.
	Using \eqref{uav_iot_sinr}, the channel capacity at time $t$ is defined as,
	\begin{equation}\label{iot_uplink}
		\begin{aligned}
			r_{i,p}^{u}(t)=
			\left\{
			\begin{array}{ll}
				\frac{\beta_{u}}{|\mathcal{I}|} \cdot \log \big(1 + \gamma_{i,p}^{u}(t)\big),\;\;\; \text{if\;$\gamma_{i,p}^{u}(t) > \gamma_{th}$}, \\
				0,\;\;\;\text{otherwise}.
			\end{array}
			\right.
		\end{aligned}
	\end{equation}
	Here $\beta_{u}$ is the fixed non-mmWave uplink channel bandwidth that is equally distributed to the IoT devices $\mathcal{I}$ at the trajectory point  $p \in \mathcal{P}$, $\gamma_{th}$ is the SINR threshold for ensuring successful uplink transmission between IoT devices and UAV-BSs. 

	\subsection{UAV-BS-to-BS Communication Model}
	The received power of the ground BS $b$ from UAV-BS $u \in \mathcal{U}$ is calculated as \cite{lai2019data},
	\begin{equation}\label{bs_power}
		\hat{P}_{b,u} = P_{b,u}^{tx} \cdot G_{u}^{tx} \cdot G_{b}^{rx} \Big(\frac{c}{4 \pi \delta_{b,u} f_{c}^{mmWave}}\Big).
	\end{equation}
	Here $P_{b,u}^{tx}$ is the transmit power of UAV-BS $u \in \mathcal{U}$ to BS $b$, $\delta_{b,u}$ is the distance between the UAV-BS $u \in \mathcal{U}$ and ground BS $b$, $c$ is the speed of light, $f_{c}^{mmWave}$ is the carrier frequency of the mmWave back-haul link, $G_{u}^{tx}$ and $G_{b}^{rx}$ are the antenna gains of the transmitter UAV $u \in \mathcal{U}$ and receiver ground BS $b$, receptively.
	At time $t$, the back-haul capacity of the channel between UAV-BS $u \in \mathcal{U}$ and ground BS $b$ at time slot $t$ is calculated as,
	\begin{equation}\label{backhaul}
		\begin{aligned}
			r_{b,u}^{mmWave}(t)=
			\left\{
			\begin{array}{ll}
				\beta_{b,u}^{mmWave} \cdot \log \big(1 + \frac{\hat{P}_{b,u}}{\beta_{b,u}^{mmWave}\sigma^{2}}\big),\text{if\;$ \delta_{u,b} \leq \bar{\alpha}$}, \\
				0,\;\;\;\text{otherwise}.
			\end{array}
			\right.
		\end{aligned}
	\end{equation}
	Here $\beta_{b,u}^{mmWave}$ is the mmWave back-haul bandwidth and $\sigma^{2}$ is the additive noise.
	If the distance $\delta_{u,b}= \sqrt{(x_{u}-x_{b})^{2} + (y_{u}-y_{b})^{2}}$ between the UAV-BS and the ground BS $b$ is less than a thresh-hold distance $\bar{\alpha}$, the UAV-BS transmits the information update to the ground base station using $\beta_{b,u}^{mmWave}$.
	Using \eqref{backhaul}, the transmission energy of UAV-BS $u \in \mathcal{U}$ while using back-haul link at time $t$ is calculated as,
	\begin{equation}\label{uav_backhaul_energy}
		E_{u}^{mmWave}(t) = P_{b,u}^{tx} \times r_{b,u}^{mmWave}(t).
	\end{equation}

	\subsection{UAV-BS Relay Network Energy Efficiency Metric Design}
	UAV-BS $u \in \mathcal{U}$ covers the observation area horizontally at a constance altitude $h_{u}$ where different UAV-BSs may maintain different altitudes.
	The assumption is practical for UAV-BSs according to the Federal Aviation Administration (FAA) regulations for small unmanned aircraft (UAS) operations.
	Moreover, the UAV-BS trajectory at time $t$ is defined as, $\tau_{u}(t) = \big[x_{u}(t), y_{u}(t)\big]^{T} \in \mathbb{R}^{2 \times 1}$.
	Therefore, the time varying distance covered by UAV-BS $u \in \mathcal{U}$ horizontally at constant altitude $h_{u}$ is defined as \cite{zeng2017energy},
	\begin{equation}\label{distance_uav}
		\delta_{u}(t) = \sqrt{h_{u}^{2}+||\tau_{u}(t)^{2}||}, 0\leq t \leq T.
	\end{equation}
	The total mobility energy cost of UAV-BS $u \in \mathcal{U}$ for covering distance $\delta_{u}$ at time $t$ is calculated as,
	\begin{equation}\label{energy_cost}
		E_{u}(t) = \delta_{u}(t) \times E_{prop}.
	\end{equation}
	Here $E_{prop} = k_{1} || v||^{3} + \frac{k_{2}}{||v||} \big(1+\frac{||a||^{2}}{g^{2}}\big)$ is the upper bound of the propulsion power consumption where $k_{1}$ and $k_{2}$ depends on the UAV-BS design and $g = \text{9.8} \;m/s^{2}$ is the gravitational acceleration \cite{eom2018uav}.
	
	%
	
	The total energy efficiency for UAV-BS $u \in \mathcal{U}$ that covers trajectory points $\mathcal{P}$ to serve IoT devices in $\mathcal{I}$ over time $T$ is defined as,
	\begin{equation}\label{energy_efficiency}
		\eta(\mathcal{P},u) = \sum_{t=1}^{T}\sum_{p = 1}^{|\mathcal{P}|}\frac{(r_{b,u}^{mmWave}(t)+ \sum_{i = 1}^{|\mathcal{I}|}r_{i,p}^{u}(t))}{(E_{u}^{mmWave}(t)+E_{u}(t))}.
	\end{equation}

	\subsection{Age of Information Model for Ground Station}
	The AoI metric is used to measure the freshness of information collected by the UAV-BSs from the trajectory points in $\mathcal{P}$ where the UAV-BSs act as relay node for the IoT devices.
	Therefore, at the BS $b$, the AoI of the trajectory $p \in \mathcal{P}$ at time $t$ is calculated as,
	\begin{equation}\label{aoi}
		\Delta_{u}(\mathcal{P},t) = t - \Delta_{u}^{\prime}(p,t), \forall p \in \mathcal{P}.
	\end{equation}
	Here $\Delta_{u}^{\prime}(p,t)$ is denoted as the most recent received data packet from the trajectory point $p \in \mathcal{P}$ at the base station $b$ by the UAV-BS $u$.
	At $t=0$, we assume $\Delta_{u}(p,t)=0$ where we adopt the the just-in-time transmission policy \cite{kaul2012real}.
	The average AoI is calculated at the base station $b$ for trajectory points $\mathcal{P}$ over time slot $T$ as,
	\begin{equation}\label{average_aoi}
		\hat{\Delta}_{b}(\mathcal{P}) = \frac{1}{T |\mathcal{P}|} \sum_{t=1}^{T} \sum_{p \in \mathcal{P}} \Delta_{u}(p,t).
	\end{equation}

	\section{Problem Formulation}
	To formulate the UAV-BS navigation optimization problem under contextual constraints (i.e., trajectory, AoI, energy efficiency constraints), first, we consider each UAV-BS $u \in \mathcal{U}$ can cover only a sub-set of trajectory points in a given time window $T$.
	Moreover, we consider sub-sets for each $u \in \mathcal{U}$ comprised of trajectory points which are denoted as, $\mathcal{P}_{u} \subset \mathcal{P}, \mathcal{P}_{u} \cap \mathcal{P}_{u^{\prime}} = \emptyset$ where $ u \neq u^{\prime}$.
	As a result, the objective of energy efficient UAV-BS navigation optimization problem is to find the cooperative trajectory path configuration of the UAV-BSs that maximizes the total energy efficiency of the UAV-BSs relay network subject to the energy and AoI metric.
	Therefore, the optimization problem formulation is represented as follows,
	\begin{align}
		&\argmax_{ \{ \mathcal{P}_u \}_{u \in \mathcal{U} }} 
		\sum_{u \in \mathcal{U}}\eta(\mathcal{P}_u,u),\\ 
		&\text{subject to} \notag \\
		& \bigcap_{u \in \mathcal{U}} \mathcal{P}_u = \{b\}, \forall u \in \mathcal{U}, \\
		& \bigcup_{u \in \mathcal{U}} \mathcal{P}_u = \mathcal{P}, \forall u \in \mathcal{U},  \\
		& \eta(\mathcal{P}_u) \geq \eta_{th}, \forall u \in \mathcal{U}, \\
		&  \hat{\Delta}_{b}(\mathcal{P}_u) \leq \hat{\Delta}_{b}^{th}, \forall p \in \mathcal{P}_{u}\backslash\{b\}.
	\end{align}
	In the above formulated problem, the constraints (15)-(18) are the trajectory, energy efficiency, and AoI constraints, respectively.
	Constraint (15) indicates non-overlapping trajectories of the UAV-BSs except the ground BS trajectory point where the information update occurs.
	Constraint (16) indicates the joint trajectory configuration of the UAV-BSs where all the trajectory points are covered interdependently.
	Constraints (17) and (18) are coupled with the decision variable $\mathcal{P}_u$ where both the energy efficiency and AoI metric are the functions of $\mathcal{P}_u$.
	Constraint (17) ensures the total energy efficiency of the UAV-BSs where the communication and mobility energy should be greater than a minimum energy efficiency threshold $\eta_{th}$.
	Finally, constraint (18) ensures the average freshness of information updates by configuration $\mathcal{P}_u$ should be less than an AoI threshold $\hat{\Delta}_{b}^{th}$.
	Due to constraint (18) in problem (14), the UAV-BSs jointly navigate different trajectory points under not only the energy efficiency constraint but also the AoI constraint where the performance of the computation oriented communication applications depend on the up-to-date information update from different trajectory points.
	Therefore, problem (14) is different from the traditional energy efficiency maximization problem for UAV navigation.
	
	The decision problem in (14) can be reduced to a base problem of vertex cover problem (i.e., Maximum Clique Problem) \cite{bomze1999maximum} with the corresponding constraints (15)-(18), which is NP-Complete.
	Similar to the maximum clique problem, problem (14) is combinatorial in nature.
	Moreover, there is no known polynomial algorithm that can tell, given a solution of (14), whether it is optimal. 
	As a result, we can infer that the decision problem in (14) belongs to the same category of the problem of the vertex cover problem, which is proven to be NP-hard.
	In the next section, we solve problem (14) with the corresponding constraints (15)-(18) by using a deep Q learning technique.
	\section{Proposed Trajectory Policy Algorithm Based on Deep Q-Learning}
	To solve problem (14), we apply a deep reinforcement-learning model which is the combination of a deep neural network and a reinforcement learning algorithm.
	Specifically, the proposed DQN approach is comprised of three components: (i) a deep neural network to reduce the dimension of the state space that is used to extract the contextual features (e.g., AoI, energy consumption) necessary for UAV-BS navigation, (ii) An experience replay memory to store the state transitions that the UAV-BS agents observe, and  (iii) an reinforcement learning (RL) framework to find the best trajectory policy that achieves the objective of problem (14) with the corresponding constraints (15)-(18).  
	Unlike the state-fo-the-art method such as control methods, the DQN does not need a network dynamic model as it is model free.
	Moreover, in the proposed approach, the use of experience replay ensures stability by breaking the temporal dependency among the observations used in the training of the deep neural network.
	
	In Section V(A), first, we model the state and action space of problem (14).
	After that, in Section V(B), we model the reward and control policy based on problem (14) with the corresponding constraints.
	Finally, in Section V(C), we provide the proposed training and testing model for UAV-BS trajectory policy.
	\subsection{State and Action Space}
	The \textit{state space} for trajectory policy of the UAV-BS is a four-dimensional state space.
	At each time step $t = \{1,2,\cdots,T\}$, the state or joint observation space of the learning agents (i.e., UAV-BS) is denoted by, $\mathcal{S} = \{s_{t} = ( p_{current}^{u}, p_{end}, \eta,\Delta  )| \eta \in [0, \eta_{th}], \Delta \in [1, \hat{\Delta}_{b}^{th}]    \}$, which corresponds to the current positions of UAV-BS $u \in \mathcal{U}$ with individual heights $h_{u}$, target position, energy efficiency of UAV-BSs, and average age for the navigation optimization.
	Moreover, the trajectory position for navigation of the UAV-BS $u \in \mathcal{U}$ is comprised of $x_{u} \in [0,X_{u}]$ and $y_{u} \in [0,Y_{u}]$, where $X_{u}$ and $Y_{u}$ are the maximum coordinate of a particular geographic location.
	Furthermore, the initial position of each of the UAV-BS is randomly assigned for each trial along with the number of IoT devices.  
	The lower and upper bounds of continuous state variables $\eta$ and $\Delta$ in the state space are calibrated from the real-world trajectory data. 
	
	The \textit{action space} of the UAV-BSs is the trajectory planning each of the UAV-BS's navigation from one feasible state (i.e., position) to the next state while satisfying the trajectory and communication constraints (i.e., constraints (15)-(18) of problem (14)).
	The learning agent selects an action $a_{t}$ from the set of available actions upon state $s_{t}$ where $a_{t} \in \mathcal{A}_{s_{t}} \subset \mathcal{A}$, and $\mathcal{A} = \{a_{1},\cdots,a_{U}\} =\{ \mathcal{P}_u \}_{u \in \mathcal{U} }$ is the configurations of the UAV-BS navigation.
	\subsection{Reward and Control Policy}
	When a learning agent implements action $a_{t}$, the environment moves to a new state $s_{t+1}$ and the immediate reward $R_{t+1}$ with the transition $(s_{t}, a_{t}, s_{t+1})$ is associated and the learning agent receives the reward through feedbacking.
	In other words, at each state transition, the agent receives the \textit{immediate reward} which is used to form the trajectory control policy for navigation.
	For future usage, the control policy is used by the learning agent that maps the current state to optimal control action.
	The immediate reward is formulated by the instantaneous energy efficiency metric of the UAV-BS's and defined as follows,
	\begin{equation}\label{immediate_reward_function}
		\begin{aligned}
			R_{t}=
			\left\{
			\begin{array}{ll}
				\alpha_{1}\eta(a_{t}),\text{if\;contraints (15)-(18) of (14) are true,} \\
				-\alpha_{1}, \text{if\;contraints (15)-(17) of (14) are violated,}\\
				0,\;\;\;\text{if\;contraints (15)-(18) of (14) is violated}.
			\end{array}
			\right.
		\end{aligned}
	\end{equation}
	Here $\alpha_{1}$ is a coefficient multiplied to the energy efficiency function and also used to penalize the agent when the constraints are violated.
	
	The objective of the learning agent over $T$ time slots is, therefore, to maximize the future reward which is defined as,
	\begin{equation}\label{expected_discounted_reward}
		\hat{R}(s,a;t) = \sum_{t_{0}=0}^{T} \gamma(t_{0}) \times R_{t}(t-t_{0}),
	\end{equation}
	Here $\gamma = [0,1]$ reflects the trade-off between the importance of immediate and future rewards which, in turn, converges to the optimal control policy.
	Moreover, the reward function \eqref{expected_discounted_reward} is obtained at time $t$ after learning the current state of the UAV-BSs over the last $T$ time steps duration.
	Therefore, we define a control policy as $\pi$ for the agent where the \textit{Q}-function or the action-value function is defined as,
	\begin{equation}\label{control_policy}
		Q^{\pi}(s,a) = \hat{R}(s,a) + \gamma \sum_{s \in \mathcal{S}} P_{s,s^{\prime}} V^{\pi}(s^{\prime}),
	\end{equation}
	Here $P_{s,s^{\prime}} $ is the transition probability of the states in the environment where $s^{\prime} = s_{t+1}$, $\pi$ is the control policy, and action $a$ is enforced though the environment simulator.
	Here, the state and reward update is based on the information received by the ground BS $b$.
	Using \eqref{control_policy}, we set the goal of our model which is to obtain the best control policy $\pi^{opt}$.
	Therefore, the maximum \textit{Q}-function is defined as,
	\begin{equation}\label{max_q_function}
		Q^{\pi^{opt}}(s,a) = \mathbb{E} \big[R + \gamma \max_{a^{\prime}} Q^{\pi^{opt}}(s^{\prime},a^{\prime})|s,a\big],
	\end{equation}  
	where the \textit{discounted cumulative state function} is,
	\begin{equation}\label{discounted_opt_function}
		V^{\pi^{opt}}(s) =  \max_{a^{\prime}}\big[ Q^{\pi^{opt}}(s,a)].
	\end{equation}
	To derive the optimal control policy $\pi^{opt}$, the \textit{Q}-function is updated as,
	\begin{equation}\label{q_update_function}
		Q_{t^{\prime}}(s,a) = Q_{t}(s,a) + \psi \big( R + \gamma \big[\max_{a^{\prime}} Q_{t} (s^{\prime},a^{\prime})\big] - Q_{t}(s,a)\big),
	\end{equation}
	Here $t^{\prime} = t+1$ and $a^{\prime} = a_{t+1}$ where the \textit{Q}-function is updated using the recursive mechanism and $\psi$ is the learning rate.
	\begin{flushright}
		\begin{algorithm}[t!]
			\textbf{Step 1: Initialization}
			
			Initialize $Q(s,a;\theta), \mathcal{M}$, target DQN parameters $\theta^{-}$ and construct DQN \\
			\Indm
			\Indp
			\textbf{Step 2: Training DQN with experience replay}\\
			\For{$e = 1,\cdots,E$}{ Initialize $\mathcal{S}$\\
				\For{$t = 1,\cdots,T$}{
					Calculate the energy efficinecy metric of the UAV-BSs using \eqref{energy_efficiency} \\
					Calculate instant reward $R_{t}$ using \eqref{immediate_reward_function}\\
					Select action $a_{t}$ with given probability $\epsilon$.\\
					Observe instant reward $R_{t}$ and next state $s_{t^{\prime}}$\\
					Store experience $(s_{t},s_{t^{\prime}},a_{t}, R_{t}, R_{t^{\prime}})$ in the experience replay memory  $\mathcal{M}$\\
					Randomly sample minibatch of experiences from  $\mathcal{M}$\\
					Adopt stochastic gradient descent (SGD) to train the DQN using loss function in \eqref{target_uniform_loss}\\
					Update $\theta$ and $Q (s,a;\theta)$ 
				}
				
			}
			Store the Q-network\\
			\textbf{Step 3: Testing UAV-BS trajectory policy for joint UAV-BS navigation}\\
			Load the stored Q-network of Step 1\\
			Retrieve $R_{t}$ of the UAV-BSs at time slot $t$\\
			Retrieve and select joint UAV-BS action $a_{t} = \max_{a_{t}} Q^{\pi^{opt}}(s_{t},a;\theta)$\\
			Update trajectory of UAV-BSs based on joint action index and target values of DQN

			%
			%
			%
			%
			\caption{DQN with experience replay for UAV-BS Trajectory 
				Policy Optimization for Navigation}
		\end{algorithm}
		\vspace{-0.5in}
	\end{flushright}
	\subsection{Training with Experience Replay and Testing UAV-BS Trajectory Policy}
	The proposed DQN approach learns how to optimally control the trajectory configurations of the UAV-BSs for navigation during the simulation.
	Therefore, it is vital for the simulation process to train a Q-network where 
	the target value for each trajectory observation environment state is given as,
	\begin{equation}\label{target_label}
		y = R + \gamma \max_{a^{\prime}} Q_{t}(s^{\prime}, a^{\prime}; \theta_{k}).
	\end{equation}
	Here we introduce $\theta_{k}$ which is the network weight obtained by the training during the $k^{th}$ iteration.
	Hence, using \eqref{target_label}, the \textit{loss function} of the training network is designed as,
	\begin{equation}\label{target_loss}
		\mathcal{L}(\theta) = \mathbb{E}_{(s,a)\sim \rho(.)}\bigg[y -  Q(s, a ;\theta))^{2}\bigg].
	\end{equation}
	Here $\rho(s,a)$ is the probability distribution over the sequences $s$ and actions $a$, $y$ is the target value of the training network which is derived from \eqref{target_label}, and the optimal network weights $\theta^{opt}$ are obtained by training.
	Furthermore, to enhance and stabilize the training of the DQN, we apply the mini-batch method to randomly collect examples from all the training episode steps $e_{t} = (s_{t},a_{t},R_{t}, s_{t^{\prime}})$ in a fixed size \textit{replay memory} $M_{t} = \{e_{1},\cdots,E\}$.
	As a result, one sample is used multiple times in the training that improves the data efficiency significantly.
	Therefore, using \eqref{target_label}, the loss function in \eqref{target_loss} is represented with a uniform distribution over $\mathcal{M}$ as,
	\begin{equation}\label{target_uniform_loss}
		   \begin{split}
		\mathcal{L}(\theta) = \mathbb{E}_{(s,a,r,s^{\prime})\sim U(\mathcal{M})}\bigg[( R + \gamma \max_{a^{\prime}} Q^{\pi^{opt}}(s^{\prime}, a^{\prime}; \theta^{-})\\ -  Q(s, a; \theta))^{2}\bigg].
		   \end{split}
	\end{equation}
	Here $U(\mathcal{M})$ is the uniform distribution over the experience replay memory $\mathcal{M}$ and $\theta^{-}$ is the stored weight parameters of the target DQN network.

	In step 1 of Alg. 1, we first initialize the network parameters randomly and introduce the target DQN with the same network structure as the original DQN network (lines 1-2 in Alg. 1).
	Then, at each training episode steps $e_{t}$, the energy efficiency metric is calculated using \eqref{energy_efficiency} considering the navigation and communication parameters in (\eqref{uav_path_loss_probability} -  \eqref{energy_efficiency}) (line 7 in Alg. 1).
	As a part of the $\epsilon$-greedy policy framework, in the exploration stage, the energy efficiency metric is used to calculate the reward function where the action is derived from the current DQN with the exploration probability $\epsilon$.
	The the reward function is observed considering the AoI context along with other constraints in problem (14) and transit to the next state $s_{t^{\prime}}$ where $t^{\prime} = t +1$.
	Moreover, the exploration stage enables the UAV-BSs to explores all the joint actions for achieving the better reward values that lead toward choosing the appropriate action with the highest energy efficiency.
	Subsequently, we adopt the mini-batch approach that shuffles the experience from the replay memory buffer at random to remove the correlation in the observation sequence, and thus, smoothing the changes in the observation data distribution (lines 11-12 in Alg. 1).
	To train the DQN, we adopt the stochastic gradient descent (SGD) algorithm using the training loss function and update the network parameter $\theta$, and network bias (lines 13-14, in Alg. 1).
	The training process stops when the UAV-BSs arrive at the terminal trajectory and the DQN network is finally stored for testing.
	
	In step 3 of Alg. 1, the stored UAV trajectory policy network in the training phase is used where $R_{t}$ of the UAV-BS at time slot $t$ is retrieved (lines 17-18 in Alg. 1).
	As a part of the exploitation in the testing phase, the action $a_{t}$ is selected for joint UAV-BS navigation where both the trajectory of the UAV-BSs and the target values of DQN are updated (lines 19-20 in Alg. 1).
	\section{Performance Evaluation}
	In this section, we first address the performance analysis experiment environment through various key metrics. 
	We then describe the outcomes obtained from the experiment and finally provide an in-depth discussion and key observations from the results of the simulation.
	In order to train the deep neural network (DNN), we consider a neural network architecture with two fully connected (FC) hidden layer with 100 hidden nodes.
	We also set the experience replay memory size, $\mathcal{M}= 200$. 
	The simulation results are obtained by averaging and normalizing the values over 100 episodes.
	\begin{table}[t]
		\caption{Simulation Settings} \label{tab:sim_tab}
		\begin{tabu}{| X[l] | X[c] |}
			\hline
			\centering
			\textbf{Simulation Parameters} & \textbf{Values}\\
			\hline
			\centering
			No. of UAV-BS  & $3$ \\
			\hline
			\centering
			No. of MEC-BS  & $1$ \\
			\hline
			\centering
			No. of IoT devices  & $100$\\
			\hline
			\centering
			No. of Trajectory points  & $[6,14]$\\
			\hline
			\centering
			Max UAV-BS heights  & $[140,250]$ (m)\\
			\hline
			\centering
			Maximum UAV-BS velocity and acceleration    & $100 \; \text{m/s}$ and , $5 \; m/s^{2}$ \\
			\hline
			\centering
			Radius of UAV-BS  & $300$ (m)   \\
			\hline
			\centering
			$f_{c}^{mmWave}$  & $28$ GHz  \cite{lai2019data}\\
			\hline
			\centering
			$\beta_{b,u}^{mmWave}$  & $20 * 100$ MHz\cite{lai2019data}\\
			\hline
			\centering
			$f_{c}$, $\beta_{u}$  &  $2$ GHz, $20$ MHz \cite{lai2019data}\\
			\hline
			\centering
			$P_{b,u}^{tx}$, $\sigma^{2}$, $\gamma_{th}$ & $20$ dBm \cite{lai2019data}, $-100$ dBm, $5$ dB\\
			\hline
			\centering
			For urban scenario, $\alpha, \hat{\alpha}, \epsilon, \bar{\epsilon}, k_{1}, k_{2}$ & $9.61,0.16,1,20,9.26\times10^{-4}, 2250$ \cite{lai2019data} \cite{eom2018uav}\\
			\hline
			\centering
			Normalized AoI threshold $\hat{\Delta_{b}^{th}}$, $\alpha_{1}$, & $[0.3, 0.9]$, $1$\\
			\hline
			
		\end{tabu}
	\end{table}
	
	\subsection{Experiment Setting}
	For the performance evaluation of the proposed approach, we consider the simulation settings which is summarized in Table \ref{tab:sim_tab}.
	In addition, we compare the proposed approach with two baseline approaches which are,
	\begin{itemize}
		\item \textit{Baseline DQN:} The structure of the baseline DQN is different from the proposed DQN with experience replay approach in terms of not having the experience replay memory.
		\item \textit{Greedy:} In case of the greedy approach, at each timeslot $t$ in an episode, each UAV-BSs $u \in \mathcal{U}$ co-operatively finds the trajectory paths for navigation that may provide the maximum immediate reward.
		In addition, the approach applies penalty for violating the system constraints in order to make fair comparison between the proposed and the baseline DQN approaches. 
	\end{itemize}
	\begin{table}[ht]
		\caption{Effects of different discount factors over the average energy efficiency (EE) of the proposed and the baseline DQN approaches.} \label{tab:fig2}
		\begin{tabu}{| X[l] | X[c] |  X[l] |}
			\hline
			\centering
			\textbf{Discount factor $\gamma$} & \textbf{DQN with replay memory (Average EE)} & \textbf{Baseline DQN (Average EE)}\\
			\hline
			\centering
			$0.4$  & $0.483663$ & $0.487285$ \\
			\hline
			\centering
			$0.5$  & $0.492033$ & $0.486141$ \\
			\hline
			\centering
			$0.6$  & $0.489998$ & $0.480957$\\
			\hline
			\centering
			$\mathbf{0.7}$  & $\mathbf{0.4927502}$ & $\mathbf{0.489459}$ \\
			\hline
			\centering
			$0.8$  & $0.490233$ & $0.486083$\\
			\hline
			\centering
			$0.9$    & $ 0.486811$ & $0.484898$ \\
			\hline
			
		\end{tabu}
	\end{table}
	\begin{table}[ht]
		\caption{Trade-off analysis between normalized average reward and normalized average AoI for different AoI thresholds.} \label{tab:fig3}
		\begin{tabu}{| X[l] | X[c] |  X[l] |}
			\hline
			\centering
			\textbf{AoI threshold $\hat{\Delta}_{b}^{th}$} & \textbf{Normalized average reward (Proposed)} & \textbf{Normalized average AoI (Proposed)} \\
			\hline
			\centering
			$0.3$  & $0.011284$ & $0.449802$ \\
			\hline
			\centering
			$0.5$  & $0.116763$ & $0.420419$ \\
			\hline
			\centering
			$\mathbf{0.7}$  & $\mathbf{0.226587}$ & $\mathbf{0.385290}$\\
			\hline
			\centering
			$0.9$  & $1.0$ & $1.0$ \\
			\hline
		\end{tabu}
	\end{table}
	\subsection{Finding Appropriate Discount Factor and AoI Threshold }
	We present the experimental results for finding the appropriate discount factor (i.e., $\gamma$) and AoI threshold (i.e., $\hat{\Delta}_{b}^{th}$) for the proposed DQN with replay memory approach.
	Here, we first find the discount factor while considering the energy efficiency metric, and then we fix the discount factor to find the appropriate AoI threshold concerning the trade-off between the average AoI and average rewards metrics.

	In Table \ref{tab:fig2} and \ref{tab:fig3}, we evaluate the appropriate discount factor and AoI threshold considering the average energy efficiency metric and we set the number of trajectory points, $|\mathcal{P}|=14$.
	In Table \ref{tab:fig2}, we observe how the discount factor affects energy efficiency.
	More specifically, when the discount factor increases in the case of the proposed DQN with replay memory, the energy efficiency metric goes up at $1.73 \%$ at the discount factor is $\gamma=0.5$.
	On the other hand, the energy efficiency metric for the baseline DQN slightly drops by $0.23 \%$ at discount factor $\gamma=0.5$ and the trend continues until discount factor $\gamma=0.6$.
	Similarly, the energy efficiency metric for the proposed approach also slightly fluctuates at discount factor $\gamma=0.6$.
	However, up to this point, the proposed DQN with replay memory still outperforms the baseline DQN approach.
	At $\gamma=0.7$, the energy efficiency metric is at the peak for the proposed DQN with replay memory and the performance of the baseline DQN also improves significantly.
	However, with a higher value of $\gamma$, the performance of both the approach decreases drastically.
	The explanation behind such phenomenon is that with a lower discount factor value (i.e., $\gamma < 0.8)$, the future reward does not matter much and the agents take action to increase the immediate reward with fewer steps forward.
	In our case, this is not suitable for reaching toward reasonable energy efficiency.
	On the other hand, with increasing value of discount factor (i.e., $\gamma > 0.7$), the agents give more attention to the benefit that future actions may bring which captures the temporal behavior of the system.
	When the discount factor is $\gamma > 0.7$, the UAV agents care too much for the future reward that eventually leads toward neglecting the immediate reward.
	So, the performance is degraded and sometimes leads to penalty as the battery of the UAV-BSs drain quickly.
	Therefore, in our simulation, we set the discount factor $\gamma = 0.7$.
	
	In Table \ref{tab:fig3}, after fixing the discount factor at $\gamma = 0.7$, we investigate the effects of different AoI thresholds to the normalized average reward and average AoI that infer to choose the appropriate AoI threshold for finding the trajectory policy for the UAV-BSs.
	We investigate the trade-off between the two performance metrics by setting four different levels of AoI threshold which are $\hat{\Delta}_{b}^{th} = [0.3, 0.5, 0.7,0.9]$.
	At $\hat{\Delta}_{b}^{th} = 0.3$, we observe that with strict AoI threshold, the normalized average reward is at the minimum whereas the AoI is at the peak point (disregarding the normalized average AoI metric value at $\hat{\Delta}_{b}^{th} = 0.9$).
	This is normal because the strict AoI threshold leads to an increasing penalty, and therefore, the performance of the system in terms of average cumulative reward decreases. 
	As we relax the AoI threshold (i.e., $\hat{\Delta}_{b}^{th} > 0.5$), the average AoI decreases significantly and the average reward increases.
	However, at $\hat{\Delta}_{b}^{th} = 0.9$, the reward value is the maximum but at this point the effect of the AoI threshold is trivial.
	More specifically, the system at this point the UAV-BSs operate disregarding the freshness of information updates at the ground BS which is not desirable.
	Therefore, we set the acceptable AoI threshold at $\hat{\Delta}_{b}^{th} = 0.7$ to balance between the average reward and the AoI metric.  
	\subsection{Experiment Results}
	\begin{figure}
		\centering
		\centering
		\includegraphics[ scale=0.50]{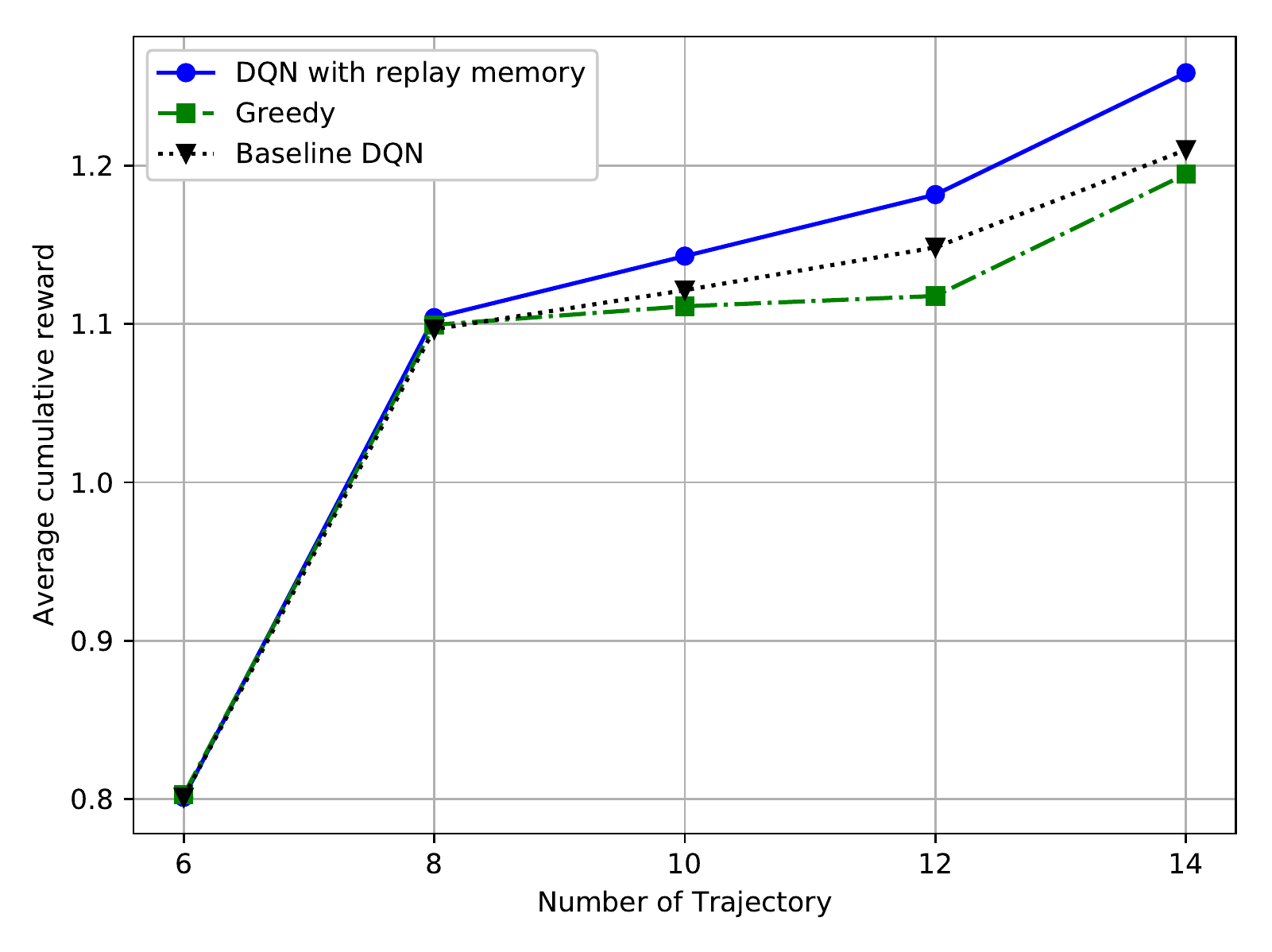}
		\caption{Average cumulative reward comparison between the proposed approach and the baseline approaches over different numbers of trajectory way-points.}
		\label{sim:fig3}
	\end{figure} 
	In Fig. \ref{sim:fig3}, we analyze the performance of the proposed DQN with a replay memory with two baseline approaches in terms of average cumulative reward with the increasing number of trajectory way-points.
	With a small number of trajectory points (i.e., up to $|\mathcal{P}|=8$), the performance of the approaches is not distinct since the density of the way-points in the geographic environment is less.
	However, as the number of trajectory points increases, the performance gaps between the approaches increase significantly.
	More specifically, the proposed DQN with the replay memory approach outperforms the greedy and baseline DQN correspondingly up to $2.84 \%$ and $1.91 \%$.
	The performance gaps between the approaches further increase with dense trajectory way-points (i.e., $|\mathcal{P}| = 14$) where the proposed DQN with replay buffer outperforms the greedy and the baseline DQN by $5.08 \%$ and $4.02 \%$, respectively. 
	
\begin{figure}
		\centering
		\centering
		\includegraphics[ scale=0.50]{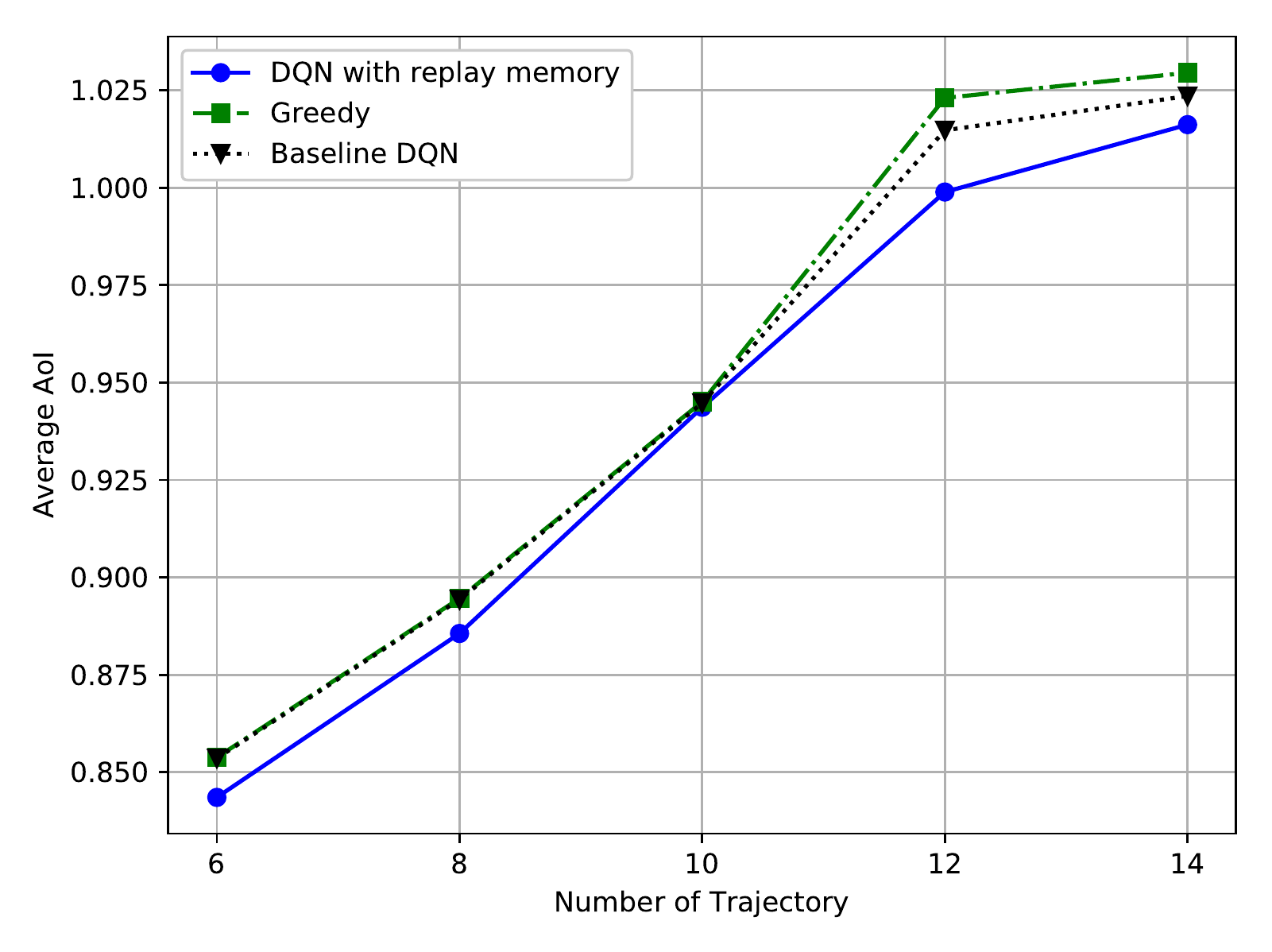}
		\caption{Average AoI comparison between the proposed approach and the baseline approaches over different number of trajectory way-points.}
		\label{sim:fig4}
	\end{figure}

	One of the key factors in the proposed model is to consider minimizing the AoI metric while finding a policy for UAV-BSs navigation.
	Therefore, in Fig. \ref{sim:fig4}, we evaluate the performance of the proposed DQN with replay memory with the greedy and baseline DQN approaches.
	As we can see from Fig. \ref{sim:fig4} that the proposed approach outperforms the greedy and baseline DQN by reducing the average AoI correspondingly up to $1.21 \%$ and $1.17 \%$.
	The performance gaps between different approaches increase gradually up to $|\mathcal{P}|=8$.
	At $|\mathcal{P}|=10$, the performance of the approaches slightly fluctuates since in the experiment we independently run the simulations with the different numbers of trajectory points.
	Therefore, this has some impacts on the experiment results.
	However, the trend of the proposed approach to reduce the AoI metric than the baseline approaches continues to carry on as the number of trajectory way-points increases.
	At $|\mathcal{P}|=14$, the performance gaps among the proposed DQN with replay memory, greedy, and baseline DQN approaches are increased by $0.72 \%$ and $1.29 \%$, respectively.
	
\begin{figure}
		\centering
		\centering
		\includegraphics[scale=0.50]{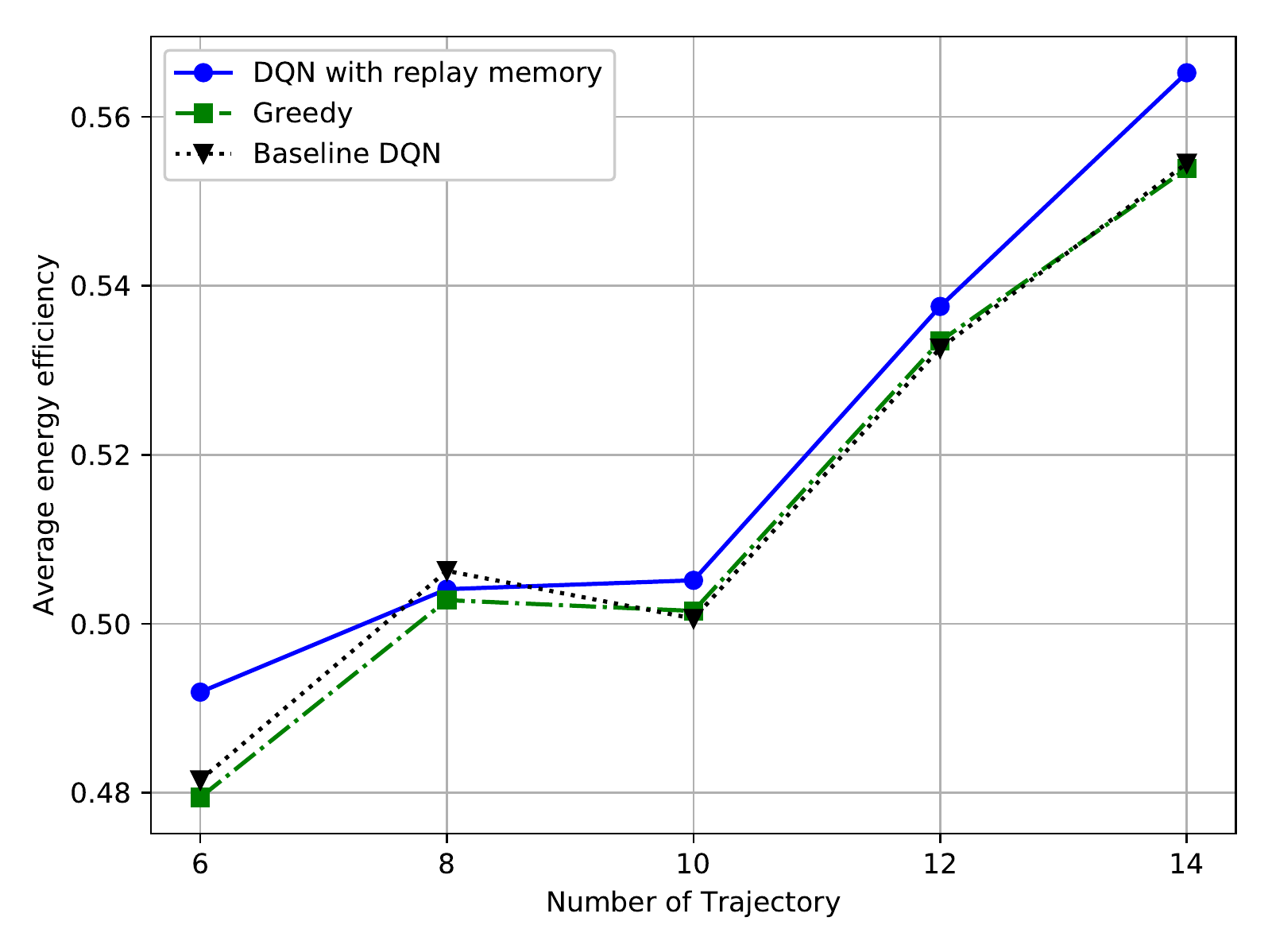}
		\caption{Average energy efficiency comparison between the proposed and the baseline approaches over different number of trajectory way-points.}
		\label{sim:fig5}
	\end{figure}

	In Fig. \ref{sim:fig5}, we show the performance of the proposed DQN with replay memory with the baseline approaches concerning the average energy efficiency.
	We observe, the proposed DQN with replay memory gradually performs better than that of the baseline approaches.
	When the number of trajectory way-points is relatively high (i.e., $|\mathcal{P}|=10$), the proposed DQN with replay memory is proven to be slightly energy efficient by $0.32 \%$ and $1.97 \%$ than the greedy and baseline DQN, respectively.
	However, the proposed approach is significantly energy efficient with the higher number of trajectory way-points (i.e., $|\mathcal{P}|=14$) where the proposed approach outperforms the greedy and baseline DQN correspondingly up to $3.6 \%$ and $3.13 \%$.   
	
\begin{figure}
		\centering
		\centering
		\includegraphics[ scale=0.50]{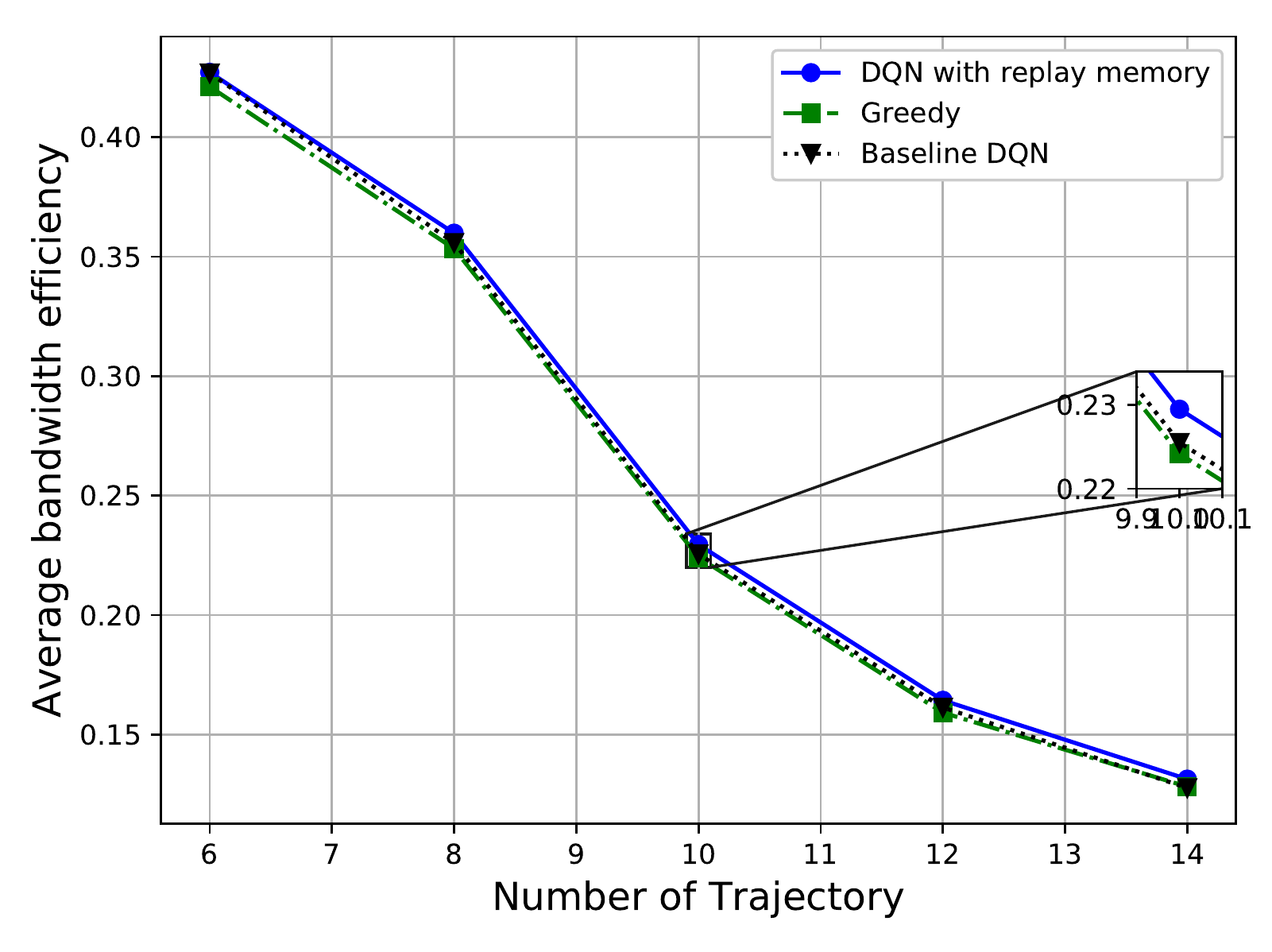}
		\caption{Average bandwidth efficiency comparison between the proposed and the baseline approaches over different number of trajectory way-points.}
		\label{sim:fig6}
	\end{figure}

	The front-haul capacity is limited with an increasing number of IoT devices at the trajectory points and therefore, the bandwidth should be utilized efficiently.
	In Fig. \ref{sim:fig6}, we compare the efficacy of the proposed approach with the greedy and baseline DQN with the varying number of trajectory points and different IoT device density.
	The proposed DQN with replay buffer efficiently utilizes the front-haul and back-haul bandwidth while traversing across different trajectory waypoints.
	The average bandwidth efficiency is quite similar when the number of trajectory points is less dense in the environment and the distance between the points is large.
	Therefore, the received interference level at the IoT devices which are served by different UAV-BS is significantly less in all the approaches.
	However, with a lightly dense trajectory way-point network with the increasing number of IoT devices (i.e., $|\mathcal{P}| =10$), all the approaches face interference and we observe a slight decrease in bandwidth efficiency.
	Nevertheless, the proposed DQN with replay buffer still outperforms the greedy and baseline DQN by $2.41 \%$ and $2.87 \%$, respectively in terms of ensuring bandwidth efficiency.   
	\begin{figure}
		\centering
		\centering
		\includegraphics[scale=0.50]{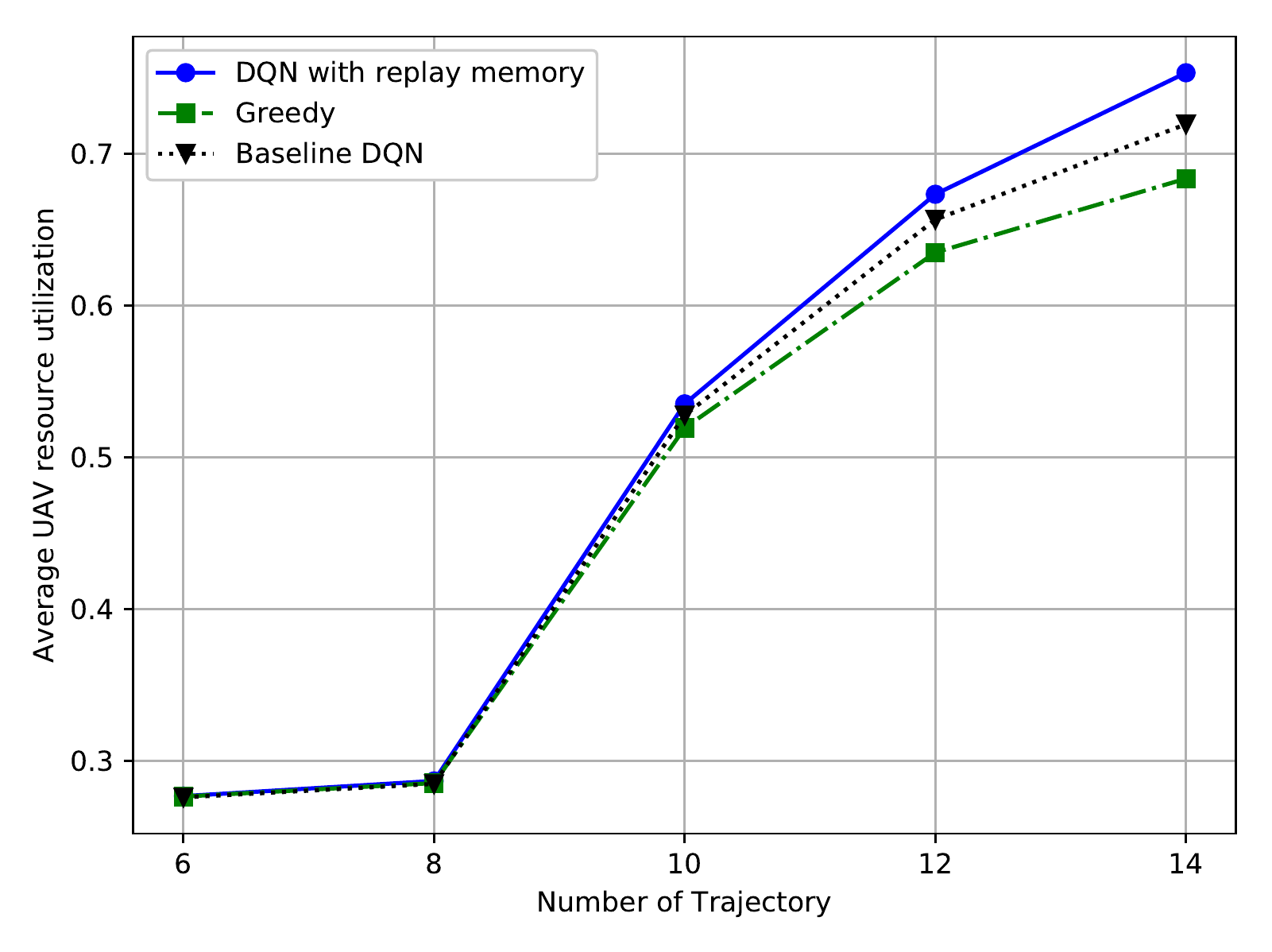}
		\caption{Average UAV resource utilization comparison between the proposed and the baseline approaches over different number of trajectory way-points.}
		\label{sim:fig7}
	\end{figure}
	
	Fig. \ref{sim:fig7} illustrates the utilization of the UAV-BSs or network resources under the proposed approach and the other two baseline approaches.
	As the number of trajectory way-points increases, the number of IoT devices using the network resources of the UAV-BSs also increases due to the increased number of associations per UAV-BSs.
	For a fixed number of UAV-BSs (i.e., $|\mathcal{U}|=3$), the IoT devices at different trajectory way-points tend to utilize the maximum network resource provided by the UAV-BSs.
	However, since the proposed DQN with replay buffer covers the trajectory way-points more efficiently than that of the baseline approaches, the network resource provided by the UAV-BSs are utilized $9.26 \%$ and  $4.71 \%$ more efficiently compared to the greedy and baseline DQN.   
	
\begin{figure}
		\centering
		\centering
		\includegraphics[ scale=0.50]{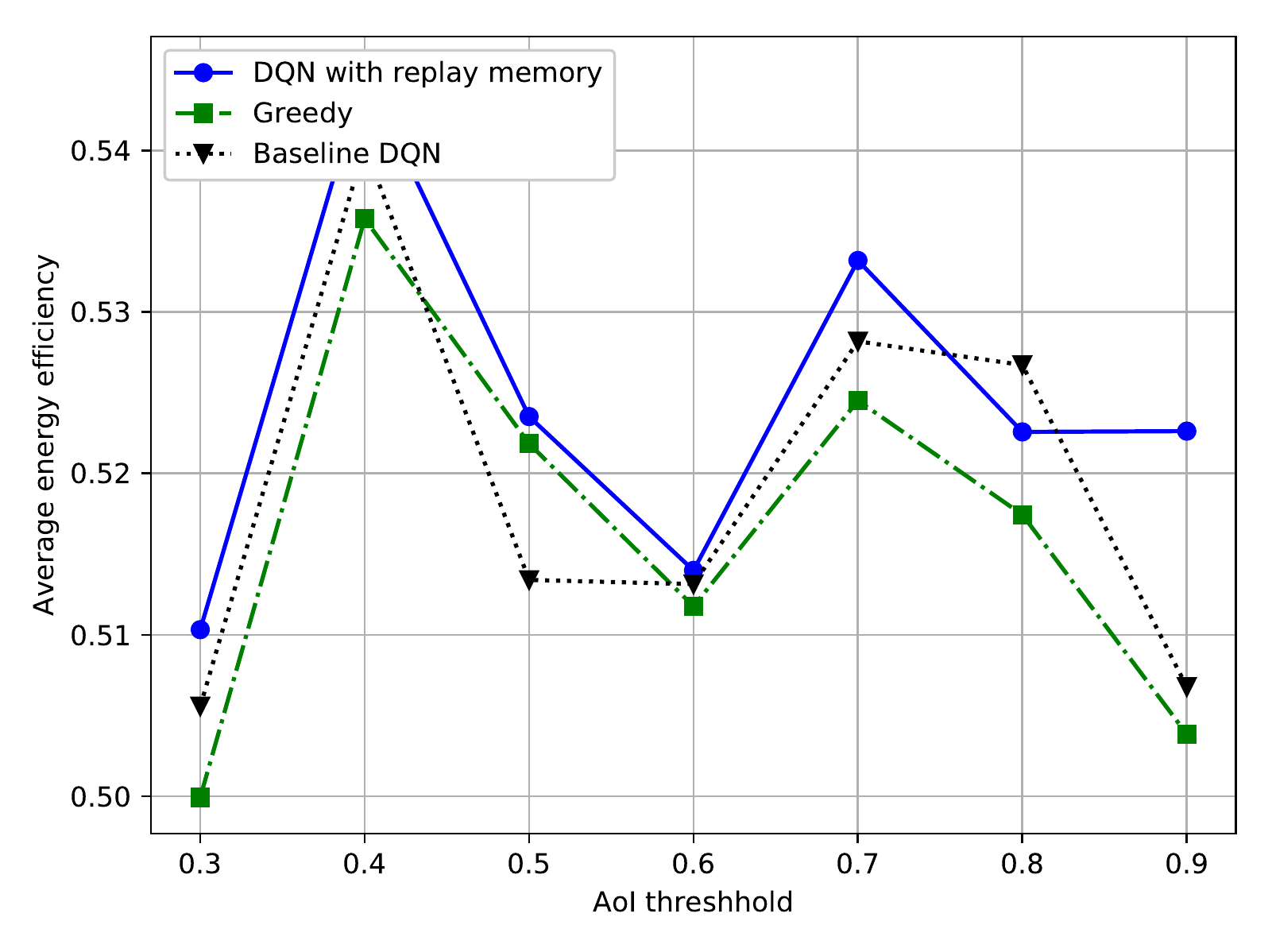}
		\caption{Average energy efficiency comparison between the proposed and the baseline approaches over different AoI threshold values.}
		\label{sim:fig8}
	\end{figure}

	Fig. \ref{sim:fig8} depicts the average energy efficiency of the UAV-BSs that operate under different AoI threshold values.
	We observe that with relatively relaxed AoI threshold $\hat{\Delta}_{b}^{th} = 0.7$ the energy efficiency of the UAV-BSs using the proposed approach is $1.63 \%$ and $0.95 \%$ more energy efficient than that of the greedy and baseline DQN, respectively.
	The performance gap between the proposed approach and the baseline DQN is relatively close because we use the same discount factor $\gamma = 0.7$.
	However, the performance gap between the proposed approach and the greedy approach is significant over all the threshold values. 
	\subsection{Discussion}
	From the experiment results described above, we find some important observations to prove the efficacy of the proposed approach than that of the baseline approaches. 
	The in-depth discussion on the experiment results can be summarized as below,
	\begin{itemize}
		\item The proposed approach can significantly enhance the UAV-BS trajectory decision where unlike the baseline DQN, the proposed approach can effectively store the transition (i.e., experience) of different environment states to reuse the transition data by random sampling.
		This stabilizes and improves the DQN training which eventually leads to better trajectory policy that considers the energy consumption of the UAV-BSs, data freshness and bandwidth utilization.
		\item We can observe from Table \ref{tab:fig2} and \ref{tab:fig3} that the objective of the UAV-BSs is largely dependent on the setting of the appropriate discount factor and the AoI threshold.
		Especially, the appropriate discount factor value can effectively enhance the training of the DQN network by not only providing better convergence but also giving a chance of improving the training of the DQN network which has a replay memory.
		\item The greedy approach sometimes performs better than that of the proposed DQN with replay memory and baseline DQN.
		However, the performance gain is limited to a small network and in the case of a large network, the proposed DQN with replay memory significantly outperforms the greedy approach.         
	\end{itemize}

	\section{Conclusion}
	In this paper, we focused on developing the UAV-BS navigation policy to improve data freshness and accessibility to the IoT network.
	As a result, we have introduced an agile deep learning reinforcement with an experience replay model that is well-suited to solving the energy-efficient UAV-BS navigation problem under trajectory and AoI constraints.
	We also performed a comprehensive simulation study to determine appropriate system parameters with the applicable discount factor and AoI efficiency metric threshold to empower the learning model.
	The simulation results show a strong correlation between energy efficiency and AoI thresholds whereby setting the proper threshold values can effectively enhance the energy efficiency and data freshness for the COC applications.
	The simulation findings also confirmed the effectiveness of the proposed DQN with experience replay memory under different network conditions.
	\ifCLASSOPTIONcaptionsoff
	\newpage
	\fi

	
	
	%
	\bibliographystyle{IEEEtran}
	\bibliography{IEEEexample}

\begin{thebibliography}{10}
\providecommand{\url}[1]{#1}
\csname url@samestyle\endcsname
\providecommand{\newblock}{\relax}
\providecommand{\bibinfo}[2]{#2}
\providecommand{\BIBentrySTDinterwordspacing}{\spaceskip=0pt\relax}
\providecommand{\BIBentryALTinterwordstretchfactor}{4}
\providecommand{\BIBentryALTinterwordspacing}{\spaceskip=\fontdimen2\font plus
\BIBentryALTinterwordstretchfactor\fontdimen3\font minus
  \fontdimen4\font\relax}
\providecommand{\BIBforeignlanguage}[2]{{%
\expandafter\ifx\csname l@#1\endcsname\relax
\typeout{** WARNING: IEEEtran.bst: No hyphenation pattern has been}%
\typeout{** loaded for the language `#1'. Using the pattern for}%
\typeout{** the default language instead.}%
\else
\language=\csname l@#1\endcsname
\fi
#2}}
\providecommand{\BIBdecl}{\relax}
\BIBdecl

\bibitem{jung2011cisco}
H.~Jung, ``Cisco visual networking index: global mobile data traffic forecast
  update 2010--2015,'' Technical Report, Cisco Systems Inc., Available online:
  https://www.cisco.com/c/en/us/solutions/collateral/service-provider/visual-networking-index-vni/white-paper-c11-738429.html,
  Tech. Rep., 2011.

\bibitem{abedin2018resource}
S.~F. Abedin, M.~G.~R. Alam, S.~A. Kazmi, N.~H. Tran, D.~Niyato, and C.~S.
  Hong, ``Resource allocation for ultra-reliable and enhanced mobile broadband
  iot applications in fog network,'' \emph{IEEE Transactions on
  Communications}, vol.~67, no.~1, pp. 489--502, 2018.

\bibitem{abedin2018fog}
S.~F. Abedin, A.~K. Bairagi, M.~S. Munir, N.~H. Tran, and C.~S. Hong, ``Fog
  load balancing for massive machine type communications: A game and transport
  theoretic approach,'' \emph{IEEE Access}, vol.~7, pp. 4204--4218, 2018.

\bibitem{zeng2018cellular}
Y.~Zeng, J.~Lyu, and R.~Zhang, ``Cellular-connected uav: Potential, challenges,
  and promising technologies,'' \emph{IEEE Wireless Communications}, vol.~26,
  no.~1, pp. 120--127, 2018.

\bibitem{8642832}
M.~S. {Munir}, S.~F. {Abedin}, N.~H. {Tran}, and C.~S. {Hong}, ``When edge
  computing meets microgrid: A deep reinforcement learning approach,''
  \emph{IEEE Internet of Things Journal}, vol.~6, no.~5, pp. 7360--7374, Oct
  2019.

\bibitem{sukhmani2018edge}
S.~Sukhmani, M.~Sadeghi, M.~Erol-Kantarci, and A.~El~Saddik, ``Edge caching and
  computing in 5g for mobile ar/vr and tactile internet,'' \emph{IEEE
  MultiMedia}, vol.~26, no.~1, pp. 21--30, 2018.

\bibitem{liu2019uav}
Y.~Liu, C.~Zhu, X.~Deng, P.~Guan, Z.~Wan, J.~Luo, E.~Liu, and H.~Zhang,
  ``Uav-aided urban target tracking system based on edge computing,''
  \emph{arXiv preprint arXiv:1902.00837}, 2019.

\bibitem{letaief2019roadmap}
K.~B. Letaief, W.~Chen, Y.~Shi, J.~Zhang, and Y.-J.~A. Zhang, ``The roadmap to
  6g--ai empowered wireless networks,'' \emph{arXiv preprint arXiv:1904.11686},
  2019.

\bibitem{8892984}
M.~S. {Munir}, S.~F. {Abedin}, and C.~S. {Hong}, ``Artificial
  intelligence-based service aggregation for mobile-agent in edge computing,''
  in \emph{2019 20th Asia-Pacific Network Operations and Management Symposium
  (APNOMS)}, Sep. 2019, pp. 1--6.

\bibitem{liu2018age}
J.~Liu, X.~Wang, B.~Bai, and H.~Dai, ``Age-optimal trajectory planning for
  uav-assisted data collection,'' in \emph{IEEE INFOCOM 2018-IEEE Conference on
  Computer Communications Workshops (INFOCOM WKSHPS)}.\hskip 1em plus 0.5em
  minus 0.4em\relax IEEE, 2018, pp. 553--558.

\bibitem{pham2018autonomous}
H.~X. Pham, H.~M. La, D.~Feil-Seifer, and L.~V. Nguyen, ``Autonomous uav
  navigation using reinforcement learning,'' \emph{arXiv preprint
  arXiv:1801.05086}, 2018.

\bibitem{pham2018cooperative}
H.~X. Pham, H.~M. La, D.~Feil-Seifer, and A.~Nefian, ``Cooperative and
  distributed reinforcement learning of drones for field coverage,''
  \emph{arXiv preprint arXiv:1803.07250}, 2018.

\bibitem{sun2016path}
Z.~Sun, J.~Wu, J.~Yang, Y.~Huang, C.~Li, and D.~Li, ``Path planning for geo-uav
  bistatic sar using constrained adaptive multiobjective differential
  evolution,'' \emph{IEEE Transactions on Geoscience and Remote Sensing},
  vol.~54, no.~11, pp. 6444--6457, 2016.

\bibitem{walker2019deep}
O.~Walker, F.~Vanegas, F.~Gonzalez, and S.~Koenig, ``A deep reinforcement
  learning framework for uav navigation in indoor environments,'' in \emph{2019
  IEEE Aerospace Conference}.\hskip 1em plus 0.5em minus 0.4em\relax IEEE,
  2019, pp. 1--14.

\bibitem{zhou2015multi}
Y.~Zhou, N.~Cheng, N.~Lu, and X.~S. Shen, ``Multi-uav-aided networks:
  Aerial-ground cooperative vehicular networking architecture,'' \emph{IEEE
  vehicular technology magazine}, vol.~10, no.~4, pp. 36--44, 2015.

\bibitem{ahmed2016energy}
S.~Ahmed, A.~Mohamed, K.~Harras, M.~Kholief, and S.~Mesbah, ``Energy efficient
  path planning techniques for uav-based systems with space discretization,''
  in \emph{2016 IEEE Wireless Communications and Networking Conference}.\hskip
  1em plus 0.5em minus 0.4em\relax IEEE, 2016, pp. 1--6.

\bibitem{7057878}
S.~F. {Abedin}, M.~G.~R. {Alam}, R.~{Haw}, and C.~S. {Hong}, ``A system model
  for energy efficient green-iot network,'' in \emph{2015 International
  Conference on Information Networking (ICOIN)}, Jan 2015, pp. 177--182.

\bibitem{wang2019autonomous}
C.~Wang, J.~Wang, Y.~Shen, and X.~Zhang, ``Autonomous navigation of uavs in
  large-scale complex environments: A deep reinforcement learning approach,''
  \emph{IEEE Transactions on Vehicular Technology}, vol.~68, no.~3, pp.
  2124--2136, 2019.

\bibitem{huang2019deep}
H.~Huang, Y.~Yang, H.~Wang, Z.~Ding, H.~Sari, and F.~Adachi, ``Deep
  reinforcement learning for uav navigation through massive mimo technique,''
  \emph{IEEE Transactions on Vehicular Technology}, 2019.

\bibitem{ragothaman2019multipath}
S.~Ragothaman, M.~Maaref, and Z.~M. Kassas, ``Multipath-optimal uav trajectory
  planning for urban uav navigation with cellular signals,'' in \emph{2019 IEEE
  90th Vehicular Technology Conference (VTC2019-Fall)}.\hskip 1em plus 0.5em
  minus 0.4em\relax IEEE, 2019, pp. 1--6.

\bibitem{zeng2019navigation}
J.~Zeng, R.~Ju, L.~Qin, Y.~Hu, Q.~Yin, and C.~Hu, ``Navigation in unknown
  dynamic environments based on deep reinforcement learning,'' \emph{Sensors},
  vol.~19, no.~18, p. 3837, 2019.

\bibitem{hu2019reinforcement}
J.~Hu, H.~Zhang, L.~Song, Z.~Han, and H.~V. Poor, ``Reinforcement learning for
  a cellular internet of uavs: protocol design, trajectory control, and
  resource management,'' \emph{arXiv preprint arXiv:1911.08771}, 2019.

\bibitem{zeng2019energy}
Y.~Zeng, J.~Xu, and R.~Zhang, ``Energy minimization for wireless communication
  with rotary-wing uav,'' \emph{IEEE Transactions on Wireless Communications},
  vol.~18, no.~4, pp. 2329--2345, 2019.

\bibitem{dong2019energy}
F.~Dong, L.~Li, Z.~Lu, Q.~Pan, and W.~Zheng, ``Energy-efficiency for fixed-wing
  uav-enabled data collection and forwarding,'' in \emph{2019 IEEE
  International Conference on Communications Workshops (ICC Workshops)}.\hskip
  1em plus 0.5em minus 0.4em\relax IEEE, 2019, pp. 1--6.

\bibitem{yang2019energy}
G.~Yang, R.~Dai, and Y.-C. Liang, ``Energy-efficient uav backscatter
  communication with joint trajectory and resource optimization,'' in \emph{ICC
  2019-2019 IEEE International Conference on Communications (ICC)}.\hskip 1em
  plus 0.5em minus 0.4em\relax IEEE, 2019, pp. 1--6.

\bibitem{hu2019uav}
X.~Hu, K.-K. Wong, K.~Yang, and Z.~Zheng, ``Uav-assisted relaying and edge
  computing: Scheduling and trajectory optimization,'' \emph{IEEE Transactions
  on Wireless Communications}, vol.~18, no.~10, pp. 4738--4752, 2019.

\bibitem{peng2019predictive}
H.~Peng, C.~Chen, C.-C. Lai, L.-C. Wang, and Z.~Han, ``A predictive on-demand
  placement of uav base stations using echo state network,'' in \emph{2019
  IEEE/CIC International Conference on Communications in China (ICCC)}.\hskip
  1em plus 0.5em minus 0.4em\relax IEEE, 2019, pp. 36--41.

\bibitem{jia2019age}
Z.~Jia, X.~Qin, Z.~Wang, and B.~Liu, ``Age-based path planning and data
  acquisition in uav-assisted iot networks,'' in \emph{2019 IEEE International
  Conference on Communications Workshops (ICC Workshops)}.\hskip 1em plus 0.5em
  minus 0.4em\relax IEEE, 2019, pp. 1--6.

\bibitem{abd2019deep}
M.~A. Abd-Elmagid, A.~Ferdowsi, H.~S. Dhillon, and W.~Saad, ``Deep
  reinforcement learning for minimizing age-of-information in uav-assisted
  networks,'' \emph{arXiv preprint arXiv:1905.02993}, 2019.

\bibitem{tong2019uav}
P.~Tong, J.~Liu, X.~Wang, B.~Bai, and H.~Dai, ``Uav-enabled age-optimal data
  collection in wireless sensor networks,'' in \emph{2019 IEEE International
  Conference on Communications Workshops (ICC Workshops)}.\hskip 1em plus 0.5em
  minus 0.4em\relax IEEE, 2019, pp. 1--6.

\bibitem{du2019joint}
Y.~Du, K.~Yang, K.~Wang, G.~Zhang, Y.~Zhao, and D.~Chen, ``Joint resources and
  workflow scheduling in uav-enabled wirelessly-powered mec for iot systems,''
  \emph{IEEE Transactions on Vehicular Technology}, vol.~68, no.~10, pp.
  10\,187--10\,200, 2019.

\bibitem{li2019minimizing}
W.~Li, L.~Wang, and A.~Fei, ``Minimizing packet expiration loss with path
  planning in uav-assisted data sensing,'' \emph{IEEE Wireless Communications
  Letters}, vol.~8, no.~6, pp. 1520--1523, 2019.

\bibitem{6497017}
Q.~{Ye}, B.~{Rong}, Y.~{Chen}, M.~{Al-Shalash}, C.~{Caramanis}, and J.~G.
  {Andrews}, ``User association for load balancing in heterogeneous cellular
  networks,'' \emph{IEEE Transactions on Wireless Communications}, vol.~12,
  no.~6, pp. 2706--2716, June 2013.

\bibitem{kalantari2016number}
E.~Kalantari, H.~Yanikomeroglu, and A.~Yongacoglu, ``On the number and 3d
  placement of drone base stations in wireless cellular networks,'' in
  \emph{2016 IEEE 84th Vehicular Technology Conference (VTC-Fall)}.\hskip 1em
  plus 0.5em minus 0.4em\relax IEEE, 2016, pp. 1--6.

\bibitem{al2014optimal}
A.~Al-Hourani, S.~Kandeepan, and S.~Lardner, ``Optimal lap altitude for maximum
  coverage,'' \emph{IEEE Wireless Communications Letters}, vol.~3, no.~6, pp.
  569--572, 2014.

\bibitem{lai2019data}
C.-C. Lai, L.-C. Wang, and Z.~Han, ``Data-driven 3d placement of uav base
  stations for arbitrarily distributed crowds,'' \emph{arXiv preprint
  arXiv:1909.11554}, 2019.

\bibitem{zeng2017energy}
Y.~Zeng and R.~Zhang, ``Energy-efficient uav communication with trajectory
  optimization,'' \emph{IEEE Transactions on Wireless Communications}, vol.~16,
  no.~6, pp. 3747--3760, 2017.

\bibitem{eom2018uav}
S.~Eom, H.~Lee, J.~Park, and I.~Lee, ``Uav-aided wireless communication design
  with propulsion energy constraint,'' in \emph{2018 IEEE International
  Conference on Communications (ICC)}.\hskip 1em plus 0.5em minus 0.4em\relax
  IEEE, 2018, pp. 1--6.

\bibitem{kaul2012real}
S.~Kaul, R.~Yates, and M.~Gruteser, ``Real-time status: How often should one
  update?'' in \emph{2012 Proceedings IEEE INFOCOM}.\hskip 1em plus 0.5em minus
  0.4em\relax IEEE, 2012, pp. 2731--2735.

\bibitem{bomze1999maximum}
I.~M. Bomze, M.~Budinich, P.~M. Pardalos, and M.~Pelillo, ``The maximum clique
  problem,'' in \emph{Handbook of combinatorial optimization}.\hskip 1em plus
  0.5em minus 0.4em\relax Springer, 1999, pp. 1--74.

\end{thebibliography}
	%
	%
	
	%
	
	%
	%
	%
	
	
	
	\vskip -2\baselineskip plus -1fil
	\begin{IEEEbiography}[{\includegraphics[width=1in,height=1.25in,clip,keepaspectratio]{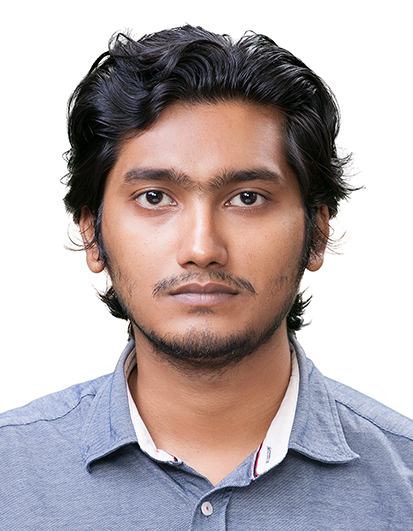}}]{Sarder Fakhrul Abedin}(S'18) received his B.S. degree in Computer
		Science from Kristianstad University, Kristianstad, Sweden, in 2013. He received  his Ph.D. degree in computer science and engineering from Kyung Hee University, South Korea in 2020. His research interests include Internet of Things (IoT) network management, Cloud computing, Fog computing, and Wireless sensor network. Mr. Abedin is a Member of the KIISE.
	\end{IEEEbiography}
	\vskip -2\baselineskip plus -1fil
	\begin{IEEEbiography}[{\includegraphics[width=1in,height=1.25in,clip,keepaspectratio]{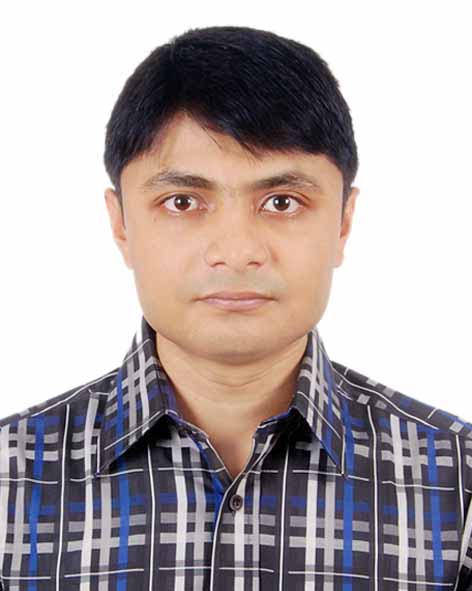}}]{Md. Shirajum Munir}(S'19) received his B.S. degree in Computer Science and Engineering from Khulna University, Bangladesh, in 2010. He served as a Lead Engineer at the solution laboratory, Samsung R\&D Institute, Bangladesh, from 2010 to 2016.  Since March 2017, he has been working toward his Ph.D. in Computer Science and Engineering at Kyung Hee University, South Korea. His research interests include the IoT network management, fog computing, mobile edge computing, software-defined networking, smart grid, and machine learning.
	\end{IEEEbiography}
	\vskip -2\baselineskip plus -1fil
	\begin{IEEEbiography}[{\includegraphics[width=1in,height=1.25in,clip,keepaspectratio]{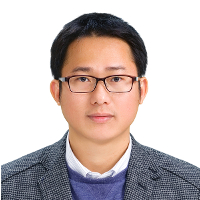}}]{Nguyen H. Tran}(S'10-M'11) received the BS degree from Hochiminh City University of Technology and Ph.D. degree from Kyung Hee University, in electrical and computer engineering, in 2005 and 2011, respectively. Since 2018, he has been with the School of Computer Science, The University of Sydney, where he is currently a Senior Lecturer. He was an Assistant Professor with Department of Computer Science and Engineering, Kyung Hee University, Korea from 2012 to 2017. His research interest is to applying analytic techniques of optimization, game theory, and stochastic modeling to cutting-edge applications such as cloud and mobileedge computing, data centers, heterogeneous	wireless networks, and big data for networks. He received the best KHU thesis award in engineering in 2011 and best paper award at IEEE ICC 2016. He has been the Editor of IEEE Transactions on Green Communications and Networking since 2016, and served as the Editor of the 2017 Newsletter of Technical Committee on Cognitive Networks on Internet of Things.
	\end{IEEEbiography}
	\vskip -2\baselineskip plus -1fil
	\begin{IEEEbiography}[{\includegraphics[width=1in,height=1.25in,clip,keepaspectratio]{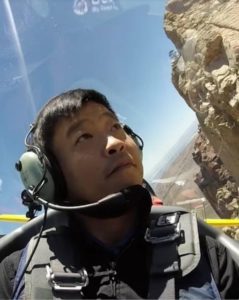}}]{Zhu~Han}
		(S'01-M'04-SM'09-F'14) received the	B.S. degree in electronic engineering from Tsinghua University, Beijing, China, in 1997, and the M.S. and Ph.D. degrees in electrical and computer engineering	from the University of Maryland, College Park, MD, USA, in 1999 and 2003, respectively. From 2000 to	2002, he was an R\&D Engineer with JDSU, Germantown, MD, USA. From 2003 to 2006, he was a Research Associate with the University of Maryland. From 2006 to 2008, he was an Assistant Professor with Boise State University, Boise, ID, USA. He is a Professor with the Electrical and Computer Engineering Department as well as with the Computer Science Department, University of Houston, Houston, TX,	USA. His research interests include wireless resource allocation and management, wireless communications and networking, game theory, big data analysis, security, and smart grids. He was the recipient of the National Science Foundation Career Award in 2010, the Fred W. Ellersick Prize of the IEEE Communication Society in 2011, the EURASIP Best Paper Award for the Journal on	Advances in Signal Processing in 2015, IEEE Leonard G. Abraham Prize in the field of Communications Systems (best paper award in IEEE JSAC) in 2016, and several Best Paper Awards in IEEE conferences. He is 1\% highly cited researcher according to Web of Science since 2017.
	\end{IEEEbiography}
	
	\vskip -2\baselineskip plus -1fil
	\begin{IEEEbiography}[{\includegraphics[width=1in,height=1.25in,clip,keepaspectratio]{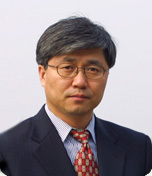}}]{Choong Seon Hong}(S'95-M'97-SM'11) received the B.S. and M.S. degrees in electronic engineering from Kyung Hee University, Seoul, South Korea, in 1983 and 1985, respectively, and the Ph.D. degree from Keio University in 1997. In 1988, he joined KT, where he was involved in broadband networks, as a Member of Technical Staff. From 1993, he joined Keio University, Japan. He was with the Telecommunications Network Laboratory, KT, as a Senior Member of Technical Staff and as the Director of the Networking Research Team until 1999. Since 1999, he has been a Professor of the Department of Computer Engineering, Kyung Hee University. His research interests include future internet, ad hoc networks, network management, and network security. He has served as a General Chair, TPC Chair/Member, or an Organizing Committee Member for international conferences, such as NOMS, IM, APNOMS, E2EMON, CCNC, ADSN, ICPP, DIM, WISA, BcN, TINA, SAINT, and ICOIN. He is currently an Associate Editor of the IEEE Transactions on Network and Service Management, International Journal of Network Management, and the IEEE Journal of Communications and Networks, and an Associate Technical Editor of the IEEE Communications Magazine. He is a member of the ACM, the IEICE, the IPSJ, the KIISE, the KICS, the KIPS, and the OSIA. 
	\end{IEEEbiography}

\end{document}